\documentclass[aps,prd,amsmath,amssymb,showpacs,showkeys,print,nofootinbib]{revtex4-1}
\usepackage{dcolumn}
\usepackage{bm}
\usepackage{hyperref}
\usepackage[retainorgcmds]{IEEEtrantools}
\usepackage[utf8]{inputenc}
\usepackage[english]{babel}
\usepackage{amsmath}
\usepackage{amsfonts}
\usepackage{amssymb}
\usepackage{graphicx}
\usepackage[caption=false]{subfig}
\usepackage{multirow}
\usepackage[normalem]{ulem}
\usepackage{xcolor}

\begin{document}

\title{2-zeroes texture and the Universal Texture Constraint}
\author{S. Gómez-Ávila} \email{selim_gomez@uaeh.edu.mx}
\author{L. López-Lozano}\email{lao_lopez@uaeh.edu.mx} \author{Pedro
  Miranda-Romagnoli} \email{pmiranda@uaeh.edu.mx}
\author{R. Noriega-Papaqui} \email{rnoriega@uaeh.edu.mx}
\affiliation{\'Area Acad\'emica de Matem\'aticas y F\'{\i}sica,
  Universidad Aut\'onoma del Estado de Hidalgo,
  Carr. Pachuca-Tulancingo Km. 4.5, C.P. 42184, Pachuca, Hgo.}
\author{Pedro Lagos-Eulogio}\email{plagos@uaeh.edu.mx}
\affiliation{Área Académica de Ingeniería Industrial, Universidad
  Autónoma del Estado de Hidalgo, Carr. Pachuca-Tulancingo Km. 4.5,
  C.P. 42184, Pachuca, Hgo.}

\begin{abstract}
  Texture matrices are a way of mitigating the redundancy inherent in
  the description of flavor physics via Yukawa couplings by
  eliminating some entries in order to identify relevant parameters. A
  four-zero texture scheme has been used in the literature to
  successfully describe fermion masses and mixing. However, as we show
  in this work, improving experimental constraints require an update
  to this analysis. In this paper the implications of a 2-zero texture
  mass matrix is studied for quarks and leptons. We show that the
  introduction of a new parameter in each mass matrix allow us to
  reach good results with relative low cost in predictability. We
  report a numerical study using a hybridized nature-inspired/cellular
  automata search algorithm. We find that leptons and quarks can be
  described by the same 1-zero structure. We describe some scenarios
  where a simplified description can be achieved, including a narrow
  region in parameter space where the same values describe charged
  leptons and neutrinos, which is a stronger version of a previously
  proposed Universal Texture Constraint.

\end{abstract}
\pacs{} \keywords{}
\maketitle

\section{Introduction}

The Standard Model (SM) provides a remarkable description of
elementary particle physics up to the sub-TeV range. One of its
shortcomings, however, is the lack of a proper understanding of
flavor; since the symmetries of the SM do not restrict the mass
matrices elements in the flavor basis, the relation between masses and
mixing angles of fermions is not understood at all. As has been
discussed by many authors \cite{GEORGI1986231}, a satisfactory
explanation of flavor can only be possible in the context of a model
with additional flavor dynamics; insofar as the mixings and masses are
described, but not explained by the SM, flavor physics \emph{is} New
Physics (NP).

The history of the attempts to face the flavor problem through texture
matrices is very long. Because the entries of the mass matrices are
not directly measured, the description of flavor through the Yukawa
couplings is characterized by redundancy and a degree of
arbitrariness; an early effort to reduce this freedom involved setting
to zero some elements of the mass matrix, called texture zeroes. Since
the pioneering paper of H. Fritzsch \cite{FRITZSCH1977436}, different
matrix-shapes (i.e., different sets of texture zeroes, or in short
\emph{textures}) have been studied in order to understand from the
\emph{bottom--up} the hierarchy between fermions masses and its
relation with discrete
symmetries. 

For the quark mass matrices, the strong hierarchy of quark masses and
the supressed off-diagonal elements of the CKM matrix have
historically suggested small mixing angles coming from radiative
corrections \cite{BALAKRISHNA1988345}, higher order terms
\cite{Verma_2013} or even the overlap of wavefunctions in extra
dimensional models \cite{strominger1985}.
On the other hand, the discovery of neutrino oscillations, which
required the addition to the SM of scalar couplings to generate
neutrino mass terms (and potentially Majorana masses) prompted the
exploration of Yukawa textures in the leptonic sector.  Since the
neutrino mixings are substantial, an approximately tribimaximal mixing
pattern was a promising avenue of research \cite{Harrison:2002er,
  Bjorken:2005rm}. Since only three angles and two mass differences
are presently known, fixing a texture gives a prediction for the
neutrino mass scale \cite{Fritzsch:2020kzp}.

If some fermion family structure survives at very high energies and
scalar bosons remain elementary, then Yukawa couplings are fundamental
constants of Nature \cite{Ibanez:1994ig}. On the other hand, if the UV
completion of the Standard Model is some GUT where quarks and leptons
(and probably some other particles) transform as a multiplet, Yukawa
couplings need also be unified, and the low-energy hierarchy could
carry an imprint of the high-energy gauge symmetry breaking pattern.

In general, the number of free parameters in a fermion mass matrix is
18, before imposing any physical constraints. Taking into account that
it is possible to make a unitary weak-basis transformation of
fermions, and that right-handed fields in the Lagrangian are singlets
under $SU(2)$, we can choose a basis where mass matrices are Hermitian
\cite{Fritzsch:1999rb}. For a pair of mass matrices, using a weak
basis transformation to make them Hermitian, and removing non-physical
phases, one ends up with ten free parameters that relate masses and
mixing angles for every sector.  Further restrictions have a
non-redundant effect in the model.

The starting point for an analysis of texture zeroes is then a 1-zero
mass matrix. In this paper we study the general properties of this
texture mass matrix for both the quark and leptonic sectors. The main
goal of this work is to establish the parameter space constrained by
the updated experimental measurements in both sectors. Also, we aim to explore the idea of common flavor structures for charged and neutral fermions; for this reason, we pay special
attention to scenarios where flavor texture universality can be
realized, like in the positive--parameters case we present in section
\ref{section6}.  

As is widely known, the neutrino sector also allows for Majorana
terms, which stem from the gauge-neutrality of sterile right-handed
neutrinos. Through the Seesaw mechanism  \cite{Minkowski:1977sc,Ramond:1979py,GellMann:1980vs,Mohapatra:1979ia}, large Majorana masses could explain the mass hierarchy between charged lepton and
neutrinos. Nevertheless, in order to simplify our analysis, in this
work we only consider neutrino masses of Dirac type. This also has the
effect of preserving the similarity (at least in the electroweak
sector) of the interaction structure of leptons and quarks.

Other choices are needed. Updated experimental results cannot
determine the sign of the $\Delta m^2_{13}$ difference, which leads to
two possibilities for the neutrino mass ordering; they are called the
normal and inverse order. In this work, we will assume normal
ordering, again with the goal of exploring a common description for
the quark and lepton sectors of the SM.  Recent experimental data,
including the Super-Kamiokande results favors Normal Order over the
inverse order \cite{Esteban:2018azc,Esteban:2020cvm}.

In order to quantify the possible common origin of the flavor
interaction in all sectors, we analyze the scaled parameters space in
the lepton sector to look for universal values that generate mass
matrices both in the charged lepton and the neutrino sector, the
so--called Universal Texture Constraint (UTC)
\cite{Carrillo-Monteverde:2020fie}.

In section \ref{section2}, the 1-zero texture matrix is defined and
its general properties are studied in the context of the SM
symmetries. In section \ref{section3} the Yukawa sector and the number
of parameters involved in the relation between masses and mixing
angles is discussed, and exact expressions for the CKM and PMNS in
terms of the 1-zero parameters are given. In section \ref{section4}, a
complete numerical analysis for the free parameters in the quark
sector is presented, using updated experimental measurements and novel
bio-inspired methods to minimize the $\chi^2$ function build with non
correlated VCKM matrix elements. In section \ref{section5} it is shown a
numerical analysis for the neutrino sector using a global fit of
experimental measurement from \cite{Esteban:2018azc} including the
theoretical restrictions and for every possible parametrization
found. Section \ref{section6} shows the region of parameters allowed
by the Universal Texture Constraint \cite{Carrillo-Monteverde:2020fie}
in the leptonic sector for real positive parameters. In section
\ref{section7}, it is given the conclusions.

\section{The 1-zero texture matrix} \label{section2}
Taking into account the invariance of the Lagrangian under a common
unitary transformation $M\to UMU^\dagger$, the mass matrix in the
flavor basis takes the general form
\begin{equation}\label{1zerotexture}
  M_F=\left(\begin{array}{ccc}
              E_F   & D_F   & 0 \\
              D^*_F & C_F   & B_F \\
              0   & B^*_F & A_F
            \end{array}\right);
          \quad F=u,d,\ell.
        \end{equation}

This is the 1-zero texture mass matrix; the zeroes at $(1,3)$ and $(3,1)$ are the only reduction of parameters that can be made by a symmetry already present in the SM Lagrangian. The matrix (\ref{1zerotexture}) has, in general, 7 parameters including phases. We write $B_F (= |B_F|\,e^{i \phi_{B_F}})$ and $D_F(= |D_F|\, e^{i \phi_{D_F}})$ in polar form, and we separate the $M_F$ matrix in the product of an orthogonal matrix and a unitary diagonal phase matrix:
        \begin{equation}\label{eq:phase_sep}
          M_F = P_F^\dagger \bar{M}_F P_F,
        \end{equation}
        where
        \begin{equation}
          P_F = \textrm{diag}\left(1, e^{i \phi_{D_F}}, e^{i(\phi_{D_F}+\phi_{B_F})}\right).
        \end{equation}
        The orthogonal matrix $\bar{M}_F$ is then written as
        \begin{equation}\label{2texture_orthogonal}
          \bar{M}_F=\left(\begin{array}{ccc}
                            E_F   & |D_F|   & 0 \\
                            |D_F| & C_F   & |B_F| \\
                            0   & |B_F| & A_F
                          \end{array}\right).
                      \end{equation}
The eigenvalues $\lambda_i^F$ of $\bar{M}_F$ are related to the fermion masses through
                      $\lambda_i^F=\pm m_i^F$; the sign depends on the
                      chosen parameterization. We diagonalize the
                      Hermitian matrix $H_F\equiv M_F M_F^\dagger$. By
                      construction, its eigenvalues are the squared
                      fermion masses
                      $\left\{(m_1^F)^2,(m_2^F)^2,(m_3^F)^2\right\}$. It
                      is possible to isolate the phases through
                      $H_F=P_F\bar{H}_F P_F^\dagger$, with
                      $\bar{H}_F= \bar{M}_F\bar{M}_F^T$ a symmetrical
                      matrix. The invariants under similarity
                      transformations are
                      \begin{align}
                        \textrm{Tr}(\bar{H}_F)& =  (m_1^F)^2+(m_2^F)^2+(m_3^F)^2,\label{invariant1}\\
                        \textrm{Det}(\bar{H}_F) & =   (m_1^F)^2(m_2^F)^2(m_3^F)^2,\label{invariant2}\\
                        \frac{1}{2}\left[\textrm{Tr}^2(\bar{H}_F)-\textrm{Tr}(\bar{H}_F^2)\right] & =  (m_1^F)^2(m_2^F)^2+(m_1^F)^2(m_3^F)^2+(m_2^F)^2(m_3^F)^2.\label{invariant3} 
                      \end{align}

                      In general, it is possible to solve this system
                      of equations for $|B_F|$, $|D_f|$ and $C_F$ in
                      terms of the parameters $E_F$ and $A_F$; for
                      comparison, we can recover the 4-zero texture
                      model by setting $E_F=0$. The system
                      (\ref{invariant1}-\ref{invariant3}) has $2^3$
                      solutions taking into account the restrictions
                      $|D_F|>0$ and $|B_F|>0$. Thus, there are $2^6$
                      ways to parameterize the mixing matrix of
                      fermions for 2-zero texture problem to determine
                      the mixing of fermions on every sector. The
                      solutions of (\ref{invariant1}-\ref{invariant3})
                      are:
                      \begin{align}
                        |B_F| & =  \sqrt{\frac{(A_F-\eta_1 m^F_1)(A_F-\eta_2 m^F_2)(A_F-\eta_3 m^F_3)}{E_F-A_F}},\label{BF}\\
                        |D_F| & =  \sqrt{\frac{(E_F-\eta_1 m^F_1)(E_F-\eta_2 m^F_2)(E_F-\eta_3 m^F_3)}{A_F-E_F}}, \label{DF}\\
                        C_F & =  -\left(A_F+E_F-\eta_1m_1^F-\eta_2m_2^F-\eta_3m_3^F\right), \label{CF}
                      \end{align}
                      where $\eta_i= \pm 1$ for $i=1,2,3$. These
                      solutions are left unchanged by
                      \begin{equation}
                        A_F \longleftrightarrow E_F \Longrightarrow  |D_F| \longleftrightarrow |B_F|,
                      \end{equation}
                      leaving invariant the element $C_F$.

                      The simplest discrete symmetry behind this
                      exchange is $S_3$, which we can represent as:
$$
[R_0]=\left(%
  \begin{array}{ccc}
    1 & 0 & 0 \\
    0 & 1 & 0\\
    0 & 0 & 1\\
  \end{array}%
\right) \quad [R_1]=\left(%
  \begin{array}{ccc}
    0 & 0 & 1 \\
    1 & 0 & 0\\
    0 & 1 & 0\\
  \end{array}%
\right) \quad [R_2]=\left(%
  \begin{array}{ccc}
    0 & 1 & 0 \\
    0 & 0 & 1\\
    1 & 0 & 0\\
  \end{array}%
\right)
$$
$$
[r_1]=\left(%
  \begin{array}{ccc}
    1 & 0 & 0 \\
    0 & 0 & 1\\
    0 & 1 & 0\\
  \end{array}%
\right) \quad [r_2]=\left(%
  \begin{array}{ccc}
    0 & 0 & 1 \\
    0 & 1 & 0\\
    1 & 0 & 0\\
  \end{array}%
\right) \quad [r_3]=\left(%
  \begin{array}{ccc}
    0 & 1 & 0 \\
    1 & 0 & 0\\
    0 & 0 & 1\\
  \end{array}%
\right).
$$

These satisfy the following properties:
\begin{itemize}
\item $[r_i]=[r_i]^{-1}=[r_i]^T.$
\item $[R_j]=[R_i]^{-1}=[R_j]^T, \ (i,j)\in (1,2).$
\end{itemize}

The action over 1-zero texture matrices is:
\begin{equation} [X_i]M_F[X_i]; \qquad X = R, r.
\end{equation}
When $[X_i] = [r_2]$ we have:
$$
[r_2]M_F[r_2]=\left(%
  \begin{array}{ccc}
    0 & 0 & 1 \\
    0 & 1 & 0 \\
    1 & 0 & 0\\
  \end{array}\right)\left(%
  \begin{array}{ccc}
    E_F   & D_F   & 0 \\
    D^*_F & C_F   & B_F \\
    0   & B^*_F & A_F
  \end{array}%
\right)\left(%
  \begin{array}{ccc}
    0 & 0 & 1 \\
    0 & 1 & 0 \\
    1 & 0 & 0\\
  \end{array}\right)=\left(%
  \begin{array}{ccc}
    A_F   & B_F   & 0 \\
    B^*_F & C_F   & D_F \\
    0   & D^*_F & E_F
  \end{array}%
\right),$$ reflecting the fact that this combination defines a family
of matrices related by an equivalence class. If we choose either
Normal Ordering of masses, i.e. $0 < m^F_1 < m^F_2 < m^F_3$, or
inverse ordering (IO) , i.e. $0 < m^F_3 < m^F_1 < m^F_2$, the viable
intervals for $E_F$ and $A_F$ are already determined for every
solution. In table (\ref{T1}), we show the different cases allowed by
the condition $A_F\geq E_F$ (NO case).

\begin{table}[h]
  \caption{\label{T1} Range of values for $A_F$ and $E_F$.}
  \begin{tabular}{|c|c|c|c|c|}
    \hline
    Case & $\eta_1$ & $\eta_2$ & $\eta_3$ & $A_F$ and $E_F$ interval  \\
    \hline \hline
    $1$ & $+1$ & $+1$ & $+1$ & $m_1^F < E_F < m_2^F < A_F < m_3^F$ \\
    \hline
    $2$ & $+1$ & $-1$ & $+1$ & $-m_2^F < E_F < m_1^F < A_F < m_3^F$ \\
    \hline
    $3$ & $-1$ & $+1$ & $+1$ & $-m_1^F < E_F < m_2^F < A_F < m_3^F$ \\
    \hline
    $4$ & $-1$ & $-1$ & $+1$ & $-m_2^F < E_F < -m_1^F < A_F < m_3^F$ \\
    \hline
    $5$ & $+1$ & $+1$ & $-1$ & $-m_3^F < E_F < m_1^F < A_F < m_2^F$ \\
    \hline
    $6$ & $+1$ & $-1$ & $-1$ & $-m_3^F < E_F < -m_2^F < A_F < m_1^F$ \\
    \hline
    $7$ & $-1$ & $+1$ & $-1$ & $-m_3^F < E_F < -m_1^F < A_F < m_2^F$ \\
    \hline
    $8$ & $-1$ & $-1$ & $-1$ & $-m_3^F < E_F < -m_2^F < A_F < -m_1^F$ \\
    \hline 
  \end{tabular}
\end{table}

As expected, to reproduce the canonical 4-zero Fritzsch texture case,
it is necessary to analyze the allowed values for $E_F$ and take the
limit $E_F \to 0$ in equations (\ref{BF}-\ref{CF}). Setting
$\eta_3=1$, we obtain:
\begin{align}
  \lim\limits_{E_F\to 0}|B_F|^2 &= \frac{(A_F - \eta_1 m_1^F)(A_F - \eta_2 m_2^F)(m_3^F -A_F)}{A_F}, \\ 
  \lim\limits_{E_F\to 0}|D_F|^2 &= -\eta_1 \, \eta_2 \frac{m_1^F \, m_2^F\, m_3^F}{A_F}, \\ 
  \lim\limits_{E_F\to 0} C_F &= -A_F +\eta_1 m_1^F +\eta_2 m_2^F + m_3^F. 
\end{align}

For the quark sector, our parameterization $2$ is reduced to the
4-zero Fritzsch texture when $\eta = -1$ in \cite{Fritzsch:2002ga} and
our parameterization $3$ is reduced to the 4-zero Fritzsch texture
when $\eta = 1$ in \cite{Fritzsch:2002ga}. When neutrinos are
introduced, our parameterization $5$ is reduced to parameterization
$1$ (see eq.11) in \cite{PhysRevD.86.053012}, our parameterization $2$
is reduced to parameterization $2$ (see eq.12) in
\cite{PhysRevD.86.053012} and our parameterization $1$ is reduced to
parameterization $1$ (see eq.13) in \cite{PhysRevD.86.053012}. The
analysis in \cite{Gupta:2013yha}, where the 1-zero texture was
previously diagonalized, corresponds to our parameterization $2$. In
the following, we proceed to analyze every possible parameterization
for the mixing matrices.


Concerning leptons, we make the following assumptions:
\begin{enumerate}
\item A Normal Ordering of the lepton masses, so that
  $|\lambda_3|>|\lambda_2|>|\lambda_1|>0$.
\item Following the idea that the elements of the mass matrix arise
  through some perturbative process, we impose a hierarchy in the
  parameters such that the $E_F < A_F$.
\item Implicitly, we are assuming a Hermitian mass matrix, so that the
  diagonalization is given by a unitary transformation\footnote{The
    general case is when
    $\text{Diag}(\lambda_1, \lambda_2, \lambda_3) = U^\dagger_L {M}
    U_R$, with $U_L\neq U_R$, is a \emph{bilinear transformation}. We
    omit this general case.}.
\item The eigenvalues of the mass matrix are related to the masses
  via
  \begin{align}
    \lambda_1 & = \eta_1 m_1,\\
    \lambda_2 & = \eta_2 m_2,\\
    \lambda_3 & =  m_3.
  \end{align}
\end{enumerate}

\section{The mixing matrices for quarks and leptons}\label{section3}

The Yukawa sector of the SM in the flavor basis has the form
\begin{equation}
  \label{eq:Yukawa_sector}
  \mathcal{L}_\text{Y}=-\bar{Q'}_L\left(Y^d\phi d'_R+Y^u\tilde{\phi}u'_R \right)-L'_LY^d\phi \ell'_R+\text{h.c.}
\end{equation}
where $Q'_L=(u'_L,d'_L)^T$ and $L'_L=(\nu'_L,\ell'_L)^T$,
$\phi=(\varphi^\pm,\varphi^0)^T$ and $\langle\varphi^0\rangle=v$. Here
$\tilde{\phi}$ is the charge conjugated of $\phi$, with its neutral
component written as $\varphi^0=v+h^0$, and the fields $d'_R$, $u'_R$
are $SU(2)$ singlets.

The existence of neutrino masses requires to go beyond the Standard
Model, adding right-handed neutrinos with Yukawa terms plus
potentially Majorana masses:
\begin{equation}
  \label{eq:Yukawa_nu}
  \mathcal{L}_{\text{Y} \nu}=- L'_L Y^\nu \phi \nu'_R + M_R \bar{\nu^C}'_R \nu'_R.
\end{equation}
In the following, we will only consider Dirac masses for the
neutrinos.

After Spontaneous Symmetry Breaking (SSB), this Lagrangian can be
separated in
\begin{equation}
  \label{mass_lagrangian}
  \mathcal{L}_{Y}=\mathcal{L}_{\text{mass}}+\mathcal{L}_{\text{CC}}+\mathcal{L}_{\text{NC}}.
\end{equation}

Changing to the mass basis through $u_{R,L}'=U^u_{R,L}u_{R,L}$,
$d_{R,L}'=U^d_{R,L}d_{R,L}$, $\ell_{R,L}'=U^\ell_{R,L}\ell_{R,L}$, and
$\ell_{R,L}'=U^\nu{R,L}\nu_{R,L}$, then omitting the contribution from
Goldstone modes, this Lagrangian becomes:
\begin{align}
  \label{eq:lagrangians}
  \mathcal{L}_\text{mass}&=-\bar d_L\bar M^dd_R-\bar u_L\bar M^uu_R-\bar \ell_L\bar M^\ell\ell_R -\bar \nu_L\bar M^\nu\nu_R +\text{h.c} \\
  \mathcal{L}_\text{NC}&=-\frac{1}{v}\left(\bar d_L\bar M^dd_R+\bar u_L\bar M^uu_R+\bar \ell_L\bar M^\ell\ell_R + \bar \nu_L\bar M^\nu\nu_R\right)h^0+\text{h.c} 
\end{align}

where
$\bar M^f= U^{F\dagger}_L(vY^F)U_R^F=\text{diag}(m_1^F,m_2^F,m_3^F)$.

The mass matrix is diagonalized through a bilinear transformation
$U_L^{f\dagger} M^fU_R^f=\text{Diag}(\lambda_1,\lambda_2,\lambda_3)$,
where $\lambda_i$ are the mass matrix eigenvalues and
$f=u,d,\ell,\nu$. In reverse, to find the matrices $U_R^f$ and $U_L^f$
we consider the diagonalization of the squared matrices
$M^fM^{f\dagger}$ and $M^{f\dagger}M^f$. These are Hermitian matrices
by definition, diagonalized by $U_f$ and $U^f_R$ respectively.  The
mixing matrices are built by the left-hand matrix $U_L^f$ through the
relations
\begin{align}
  V_\text{CKM}&=U^{u\dagger}_LU^d_L,\label{eqs:mixingmatrices1}\\
  V_\text{PMNS}&=U^{\nu\dagger}_LU^\ell_L,\label{eqs:mixingmatrices2}
\end{align}
for quarks and leptons respectively. Since the measured observables
are just the masses and mixing angles, the matrix $U_R^f$ is
non-observable. When the mass matrix is Hermitian then
$U_L^f=U_R^f$. Mixing matrices (\ref{eqs:mixingmatrices1}) and
(\ref{eqs:mixingmatrices2}) assume Dirac masses for the fermions;
additionally, the neutrino sector also allows for Majorana mass terms.

It is possible to chose a special flavor basis where the flavor
symmetries are explicit. In particular, the reduction of free
parameters needed to describe the relation between fermions masses and
mixing angles introduce significant zeroes in the mass matrices; this
is what the literature refers to as choosing a texture mass
matrix. Once the Hermitian nature of the mass matrix is established,
the position of the zeroes in this matrix determines the number of
free parameters. A zero outside the diagonal represent a reduction by
2 parameters, i.e. the norm and the phase; a zero in the diagonal only
removes one real parameter.

In order to make a predictive model from a texture mass matrix two
objectives are pursued:
\begin{itemize}

\item To reproduce the measured mixing angles through suitable
  functions depending on the measured fermion masses. It is important
  to include only the non-correlated measurements, as the mixing
  angles and the Jarlskog index or the independent mixing matrix
  elements in (\ref{eqs:mixingmatrices1}) and
  (\ref{eqs:mixingmatrices2}). In this work, we have chosen a
  parametrization proposed by M. Kobayashi and T. Maskawa as is
  described in \cite{Kobayashi:1973fv,PDG} for the case of quarks. For
  the leptonic sector the PMNS matrix was parameterized as in
  \cite{Gupta:2013yha}, where the mixing matrix is separated as the
  product of three rotation matrices and a CP violating phase
  matrix. In the quark sector, for convenience, we have taken the
  matrix elements of the first row of the CKM matrix and the Jarlskog
  index to determine allowed regions of values for the free parameters
  of the model. For the leptonic sector, we have chosen also the
  first-row PMNS matrix elements and the element $(2,3)$ that contains
  the CP phase in this sector. Both choices are equivalent because the introduction
  of a texture matrix relates these parameters to mixing angles and the
  CP phase.

\item To introduce the smallest number of free parameters. In general,
  a $3\times 3$ Hermitian matrix has 9 free parameters. If all
  eigenvalues are known, then this is reduced by 3; an additional
  restriction can be obtained from the CP violating phase through the
  Jarlskog index.  Therefore, a mass matrix has 5 free parameters
  before the introduction of a texture. In this work we have
  considered non-degenerated masses in the leptonic sector as they are
  in the quark sector. Below, we present a detailed description of the
  remaining free parameters.
\end{itemize}


To build a mixing matrix two matrices are needed, $U_L^{f_u}$ and
$U_L^{f_d}$, where $f_d$ and $f_u$ represent the down-type and up-type
fermions: $f_u=u,\nu$ and $f_d=d,\ell$. For every type of fermions a
mass matrix texture is introduced. Every Hermitian matrix has 9
parameters, so mixing matrices are expressed in terms of 18
parameters. Once we remove some parameters with 3 invariants (3
parameters), one CP violating phase (1 parameter), the mixing angles
(3 parameters) and by using the rephasing freedom, 10 parameters
remain to be fixed.

Imposing the same texture matrix for the up-type and down-type
fermions produces what is called a \emph{parallel texture}; this
duplicates the number of restrictions on the free parameters. For
instance, if a non-diagonal zero is put on a texture, the number of
free parameters is reduced by 4, leaving 6 parameters to describe the
mixing matrix. This case is studied in this work.

There are additional possibilities to further reduce the number of
free parameters; two interesting examples will be commented on.

\begin{itemize}
\item The first is the \emph{Universal Texture Constraint}, where a
  relation between different types of fermions is considered such that
  the same values used for the parameters in the up-type texture mass
  matrix are used in the down-type without taking into account the
  phases; in this way, the number of parameters is reduced to 3
  parameters and one phase.

  The quark sector is one of the most studied, and the mixing angles
  and CP phase have been measured with remarkable precision. This
  measurements strongly restrict the viable textures for the mass
  matrix. In particular, the 4-zero texture, where 2 zeroes come from
  the up-type quark sector and the rest from the down-type quarks has
  been extensively studied.

  On the leptonic sector, on the other hand, many open questions
  remain.  One is the scale of the masses of the neutrinos, and the
  Dirac or Majorana nature of such masses. Unlike the quark sector,
  where all masses are known, in the leptonic sector only the mixing
  angles and the differences of the squared masses are known, so there
  is one extra free parameter. It is possible to choose this free
  parameter as the heaviest neutrino mass ($m_{\nu_3}$), and we have
  done so.

  Because there is some freedom in the election of the flavor basis,
  in this work two cases are analyzed. First, it is assumed that the
  flavor basis where the mass matrix for neutrinos have some texture
  zeroes, and the one for the charged lepton is diagonal. In other
  words, there are not free parameters coming from the charged lepton
  sector. Nevertheless, as it was shown below, in light of the updated
  fitting from the NuFit2019 collaboration, it is no possible to
  fulfill the experimental condition on the mixing angles and the
  Jarlskog index with this small set of free parameters.

\item The second analyzed possibility is the mass matrix with texture
  zeroes for the charged lepton sector. Although there is a large
  number of possibilities to chose a texture zero with a smaller
  number of new parameters, in this work the Hermitian mass matrix
  with a non-diagonal zero in the charged lepton sector is studied,
  because it is the most general matrix that preserves all symmetries
  of the SM. Although some authors claim that this is not a very
  predictive model because it introduce many new parameters, it is
  interesting to explore the common origin of flavor analyzing the
  case of a \emph{parallel texture} between charged leptons and neutrino
  sector.

  In order to analyze how is the free parameters space of the 1-zero
  texture mass matrix, we have made the assumption that there is a
  parallel structure between charged leptons and neutrinos sectors,
  this is, the same texture for the neutrino mass matrix of the
  charged leptons is used in the neutrino sector. This duplicates the
  number of free parameters introduced in the model of the PMNS matrix
  respect the one where the charged leptons basis is the mass
  basis. Nevertheless, to reproduce the updated experimental results,
  more parameters are needed, because we can demonstrate that if we
  work in the mass basis for the charged lepton and the flavor basis
  in the neutrino sector, this set is not enough. In this analysis we
  have adopt an scheme without approximations.
\end{itemize}

Invariants of the mass matrix under a similarity transformation
produce $n$ relations between free parameters. The number of free
parameters for a Hermitian texture is given by
\begin{equation}
  \mathcal{N}(n,n_d,n_\text{nd})=n^2-n-n_\text{d}-n_\text{nd},
\end{equation} 
where $n$ is the number of generations {(number of invariants under
  unitary transformations)}, $n_d$ the number of zeroes in the
diagonal and $n_\text{nd}$ the number of parameters {reduced by the
  off-diagonal zeroes (for one off-diagnal zero $n_{nd}=2$, see
  ref. \cite{FRITZSCH20001})}. In the quark and the charged lepton
sector, the masses of fermions are known, but in the neutrino sector
only the mass differences are known and the heaviest neutrino mass is
yet to be determined. It is however possible to deduce through
cosmological observations an upper bound on this heaviest neutrino
mass.

As previously discussed, without loss of generality we can reduce the
shape of the mass matrix to the texture with one off-diagonal
zero. For three generations ($n=3$) and with fermion masses given as
known, as in the case of quarks and charged leptons, using the three
restrictions coming from the matrix invariants, and exploiting the
rephasing of the fermion fields, the number of free parameters
$\mathcal{P}$ of a Hermitian texture is given by
\begin{equation}
  \mathcal{P}=\mathcal{N}(3,0,2)-n_\alpha=3,
\end{equation} 
{where $n_{\alpha}=1$ is the number of phases removed by rephasing
  fermion fields. The mixing matrix for every sector thus has
  $2\mathcal{P}$ free parameters that have to be restricted by
  additional measurements.} Another source of information on the
parameters of the texture is the mixing angles of fermions and the CP
violating phase. Thus in the quark sector when an off-diagonal 1-zero
texture is used to model the mass matrix we have 2 free parameters to
explain the relation between the quarks masses and the mixing angles,
or equivalently the non--correlated $V_\text{CKM}$ matrix elements.

In the lepton sector, the known observables are $\Delta m^2_{21}$,
$\Delta m^2_{32}$, three mixing angles and a CP violating phase, so there is one less observable with respect to the quark sector. In
the next section we describe the issues introduced by the simplest
scheme where the mass matrix of the charged leptons is diagonal, and
discuss the phenomenological viability of the parallel texture assumption
when the updated experimental measurements on the neutrino sector are
taken into account.

Finally, the theoretical flavor mixing matrices $V_{CKM}$ and
$U_{PMNS}$ arising from 2-zero textures are :
\begin{align}
  V_{CKM}^{th} &= U^{u\dagger}_L\, U^d_L = O_u^T\, (P_u\,P_d^{\dagger})\, O_d, \nonumber \\
  \label{VCKM}
  \left( V_{CKM}^{th} \right)_{i\, \alpha} &= (O_u)_{1i}\,(O_d)_{1 \alpha} + (O_u)_{2i}\,(O_d)_{2 \alpha} \, e^{i\, \phi^q_1} + (O_u)_{3i}\,(O_d)_{3 \alpha} \, e^{i(\phi^q_1 + \phi^q_2)},  \\ 
  V_{PMNS}^{th} &= U^{\nu \dagger}_L\, U^l_L = O_{\nu}^T\, (P_{\nu}\,P_l^{\dagger})\, O_l, \nonumber \\
  \label{UPMNS}
  \left( V_{PMNS}^{th} \right)_{j\, \beta} &= (O_{\nu})_{1j}\,(O_l)_{1 \beta} + (O_{\nu})_{2j}\,(O_l)_{2 \beta} \, e^{i\, \phi^l_1} + (O_{\nu})_{3j}\,(O_l)_{3 \beta} \, e^{i(\phi^l_1 + \phi^l_2)},
\end{align}
where the phases are defined as $\phi^q_1 = \phi_{Du} - \phi_{Dd}$,
$\phi^q_2 = \phi_{Bu} - \phi_{Bd}$,
$\phi^l_1 = \phi_{D\nu} - \phi_{Dl}$,
$\phi^l_2 = \phi_{B\nu} - \phi_{Bl}$ and the $O$ matrix elements are:
\begin{align}
  \label{Os}
  (O_F)_{11} &= \left[ 1 + \frac{(E_F- \eta_1\,m_1^F)(A_F-E_F)}{(\eta_2\,m_2^F -E_F)(\eta_3\,m_3^F -E_F)} + \frac{(E_F -\eta_1\,m_1^F)(A_F- \eta_2\,m_2^F)(\eta_3\,m_3^F -A_F)}{(\eta_2\,m_2^F -E_F)(\eta_3\,m_3^F -E_F)(A- \eta_1\,m_1^F)}  \right]^{-1/2} \\
  (O_F)_{22} &= \left[ 1 + \frac{(E_F- \eta_1\,m_1^F)(\eta_3\,m_3^F-E_F)}{(\eta_2\,m_2^F -E_F)(A_F -E_F)} + \frac{(A_F -\eta_1\,m_1^F)(\eta_3\,m_3^F -A_F)}{(A_F -E_F)(A_F- \eta_2\,m_2^F)}  \right]^{-1/2} \\
  (O_f)_{33} &= \left[ 1 + \frac{(A_F-E_F)(\eta_3\,m_3^F -A_F)}{(A_F-\eta_2\,m_2^F)(A_F-\eta_1\,m_1^F)} + \frac{(E_F -\eta_1\,m_1^F)(\eta_2\,m_2^F- E_F)(\eta_3\,m_3^F -A_F)}{(\eta_3\,m_3^F -E_F)(A_F -\eta_2\,m_2^F)(A_F- \eta_1\,m_1^F)}  \right]^{-1/2}  \\
  (O_F)_{2i} &= \frac{\eta_i\,m_i^F -E_F}{|D_F|}\,(O_F)_{1i} \quad i=1,\,2,\,3. \\
  (O_F)_{3i} &= \frac{|B_F|}{\eta_i\,m_i^F -A_F} \,(O_F)_{2i} \quad i=1,\,2,\,3.  
\end{align}

\section{Numerical analysis of the 2 zero texture in the quark
  sector}\label{section4}

Equation (\ref{VCKM}), gives the theoretical quark flavour mixing
matrix $V_{CKM}^{th}$ as an explicit function of texture quark model
parameters $A_u$, $A_d$, $E_u$, $E_d$, $\phi^q_1$ and $\phi^q_2$. In
order to find a numerical range for these parameters according with
the experimental $V_{CKM}$ values a $\chi^2$ minimization procedure is performed. Because they are related by the unitary conditions
  ($V_{CKM}\,V_{CKM}^{\dagger}=I$), not all $V_{CKM}$ elements are
  independent. We can choose the four independent parameters as follows: three mixing angles and one CP-violating complex
  phase. Following~\cite{Felix-Beltran:2013tra,Barranco:2010we}, we
  take $\left| V_{ud} \right|$, $\left| V_{us} \right|$,
  $\left| V_{ud} \right|$ and the Jarlskog invariant as experimental parameters for the $\chi^2$ analysis.   We define:

{\small \begin{equation}
    \chi^{2}(A_u,\,A_d,\,E_u,\,E_d,\,\phi^q_1,\,\phi^q_2) = \frac{
      \left( \left| V_{ud}^{^{th}} \right| - \left| V_{ud} \right|
      \right)^{2} }{ \sigma_{V_{ud}}^{2} } + \frac{ \left( \left|
          V_{us}^{^{th}} \right| - \left| V_{us} \right| \right)^{2}
    }{ \sigma_{V_{us}}^{2} } + \frac{ \left( \left| V_{ub}^{^{th}}
        \right| - \left| V_{ub} \right| \right)^{2} }{
      \sigma_{V_{ub}}^{2} } + \frac{ \left(\mathcal{J}^{^{th}}_q -
        \mathcal{J}_q \right)^2}{\sigma_{{\mathcal{J}_q}}^2},
    \label{Chi_quarks}
  \end{equation}}
where the terms with super-index $``th"$ are given in (\ref{VCKM}) and the quantities without super-index are the experimental data with uncertainty $\sigma_{V_{kl}}^{2}$ taken from \cite{PDG}.

Since  we have 4 experimental observables, as given in the above equation (\ref{Chi_quarks}), ($N_{obs} = 4$), a good criteria is to find the
numerical range of parameters $A_u$, $A_d$, $E_u$, $E_d$, $\phi^q_1$
and $\phi^q_2$, that gives a value of $\frac{\chi^{2}}{N_{obs}}$ less
or equal to $1$.
\begin{equation}
  \label{OFuntion}
  \frac{\chi^{2}(A_u,\,A_d,\,E_u,\,E_d,\,\phi^q_1,\,\phi^q_2)}{N_{obs}} \leq 1
\end{equation}

In this analysis, we will consider that u-quark mass matrix and
d-quarks mass matrix have the same parameterization (see table
\ref{T1}). This automatically gives us eight working scenarios (see
table \ref{T2}), which we will analyze numerically.

\begin{table}[htbp]
  \begin{tabular}{|c|c|c|c|c|c|c|}
    \hline
    Scenario & $A_u$ & $A_d$ & $E_u$ & $E_d$ & $\phi^q_1$ & $\phi^q_2$ \\
    \hline \hline
    $1$ & $(m_c,\, m_t)$ & $(m_s,\, m_b)$ & $(m_u,\, m_c)$ & $(m_d,\, m_s)$ & $(0,\, 2\pi)$ & $(0,\, 2\pi)$ \\
    \hline
    $2$ & $(m_u,\, m_t)$ & $(m_d,\, m_b)$ & $(-m_c,\, m_u)$ & $(-m_s,\, m_d)$ & $(0,\, 2\pi)$ & $(0,\, 2\pi)$ \\
    \hline
    $3$ & $(m_c,\, m_t)$ & $(m_s,\, m_b)$ & $(-m_u,\, m_c)$ & $(-m_d,\, m_s)$ & $(0,\, 2\pi)$ & $(0,\, 2\pi)$ \\
    \hline
    $4$ & $(-m_u,\, m_t)$ & $(-m_d,\, m_b)$ & $(-m_c,\, -m_u)$ & $(-m_s,\, -m_d)$ & $(0,\, 2\pi)$ & $(0,\, 2\pi)$ \\
    \hline
    $5$ & $(m_u,\, m_c)$ & $(m_d,\, m_s)$ & $(-m_t,\, m_u)$ & $(-m_b,\, m_d)$ & $(0,\, 2\pi)$ & $(0,\, 2\pi)$ \\
    \hline
    $6$ & $(-m_c,\, m_u)$ & $(-m_s,\, m_b)$ & $(-m_t,\, -m_c)$ & $(-m_b,\, -m_s)$ & $(0,\, 2\pi)$ & $(0,\, 2\pi)$ \\
    \hline
    $7$ & $(-m_u,\, m_c)$ & $(-m_d,\, m_s)$ & $(-m_t,\, -m_u)$ & $(-m_b,\, -m_d)$ & $(0,\, 2\pi)$ & $(0,\, 2\pi)$ \\
    \hline
    $8$ & $(-m_c,\, -m_u)$ & $(-m_s,\, -m_d)$ & $(-m_t,\, -m_c)$ & $(-m_b,\, -m_s)$ & $(0,\, 2\pi)$ & $(0,\, 2\pi)$ \\
    \hline
  \end{tabular}
  \caption[] {\label{T2} Scenarios of quark analysis.}
\end{table}

To search the set of parameters that satisfy the cost function
criteria (\ref{OFuntion}), it is necessary to implement a specific
search technique and, in this area, it is common to use exhaustive
search techniques which work very well for optimization of small
number of parameters (6-textures, 5-textures, 4-textures) and assuming
certain simplifying conditions \cite{Gupta:2009ur,Verma:2010jy}. On
the other hand, if we implement traditional optimization techniques
like the gradient method, Davidon-Fletcher-Powell method or Newton
method, great difficulties ensue because the nonlinear dependence
among the quantities ($O$ matrix elements) in general leads to convergence
problems.

In the last two decades, meta-heuristic algorithms have been applied
to many applications due to their efficiency, reliability and
relatively low computation time \cite{azizyan}. There is a great
diversity of heuristic algorithms, which imitate strategies developed
by animals and plants for their survival. One of the most important
algorithms is the Particle Swarm Optimization (PSO)\cite{Kennedy1995},
due to its relatively ease of implementation and good convergence
speed, although the particles are easily trapped by local
optima. There are many variants of the PSO that improve its
performance \cite{freitas}. The Differential Evolution algorithm (DE)
\cite{PriceStorn1997a} is another widely used heuristic due to its
robustness \cite{das}, although the speed of convergence is quite
low. To improve the two main disadvantages of PSO and DE, various
hybrid methods have been presented, such as CPSO-DE \cite{lagos2017},
that combines the nature inspired PSO with a cellular automata
\cite{ShiLiuGaoEtAl2011}, and an evolution rule based on DE. In this
paper, we use a variant of CPSO-DE that uses a mutation probability to
create a mutation/donor vector using the current best individual found
so far ($DE/best/2$) and a random vector ($DE/rand/1$). A methodology
based on the optimization of constraints is also used.

In our numerical algorithm, we use the following quark masses values
\cite{PDG}:
$$
\begin{array}{ccc}
  m_u &=& 2.16^{+0.49}_{-0.26}\, MeV  \\
  m_c &=& 1.27^{+0.02}_{-0.02}\, GeV  \\
  m_t &=& 172.76^{+0.30}_{-0.30}\, GeV 
\end{array}
\qquad
\begin{array}{ccc}
  m_d &=& 4.67^{+0.48}_{-0.17}\, MeV \nonumber \\
  m_s &=& 93^{+11}_{-5}\, MeV \nonumber \\
  m_b &=& 4.18^{+0.03}_{-0.02}\, GeV \nonumber
\end{array}
$$

And the $V_{CKM}$ current values \cite{PDG}:
$$
\begin{small}
  V_{CKM}= \begin{pmatrix}
    0.97401 \pm 0.00011 & 0.22650 \pm 0.00048 & 0.00361^{+0.00011}_{-0.00009} \\
    0.22636 \pm 0.00048 & 0.97320 \pm 0.00011 & 0.04053^{+0.00083}_{-0.00061} \\
    0.00854_{-0.00016}^{+0.00023} & 0.03978^{+0.00082}_{-0.00060} &
    0.999172^{+0.000024}_{-0.000035}
  \end{pmatrix}
\end{small}
$$
and the Jarlskog invariant value
$J_q = (3.00^{+0.15}_{-0.09}) \times 10^{-5}$.

\begin{table}[htpb]
  \begin{center}
    \begin{tabular}{ | c | c | m{9cm} |}
      \hline Scenario & $\chi^2$ Minimum & Quarks Parameters Value \\ \hline 
      \hline
      1 & $0.00125$ &$A_u = 5.382829 \times 10^{4} \, MeV$, $A_d =1.225015 \times 10^{3}\, MeV$, $E_u =6.995365\times 10^{1}\, MeV$, $E_d=2.177576\times 10^{1}\, MeV$, $\phi^q_1 =6.045549$, $\phi^q_2 =6.236893$. \\ \hline 
      2 & $0.0025$ & $A_u =9.416563\times 10^2\, MeV$, $A_d =7.763994\times 10^1\, MeV$ $E_u=-1.648069\times 10^2\, MeV$, $E_d=2.666386\, MeV$, $\phi^q_1=2.829067$,  $\phi^q_2 =5.350638$. \\ \hline
      3 & $0.00047$ & $A_u =1.562226\times 10^4\, MeV$, $A_d =4.194742\times 10^2\, MeV$, $E_u =7.206153\times 10^1\, MeV$, $E_d =1.417096\, MeV$, $\phi^q_1 =5.217658$, $\phi^q_2 =6.168979$. \\ \hline
      4 & $0.00389$ & $A_u =1.726981\times 10^5\, MeV$, $A_d =6.790165\times 10^1\, MeV$, $E_u =-4.156814\, MeV$, $E_d =-9.485004\, MeV$, $\phi^q_1 =3.166315$, $\phi^q_2 =5.648380$. \\ \hline
      5 & $0.0025$ & $A_u =5.646824\times 10^1\, MeV$, $A_d =1.824668\times 10^1\, MeV$, $E_u =-1.055435\times 10^5\, MeV$, $E_d=-2.680576\times 10^3\, MeV$, $\phi^q_1 =3.114647$, $\phi^q_2 =3.572330$. \\ \hline
      6 & $1.38\times 10^{5}$ &  \\ \hline
      7 & $0.0025$ & $A_u =5.519055\times 10^1\, MeV$, $A_d =1.214369\times 10^1\, MeV$, $E_u =-4.582454\times 10^4\, MeV$, $E_d =-1.107061\times 10^3\, MeV$, $\phi^q_1 =3.086221$, $\phi^q_2 =4.077284$. \\ \hline
      8 & $1.38\times 10^{5}$ &  \\ \hline  
    \end{tabular}
  \end{center}
  \caption[] {\label{tdatos} $\chi^2$ minimum and their
    corresponding numerical values of the parameters.}
\end{table}

\begin{figure}[htbp]
  \begin{center}
    \includegraphics[scale=0.65]{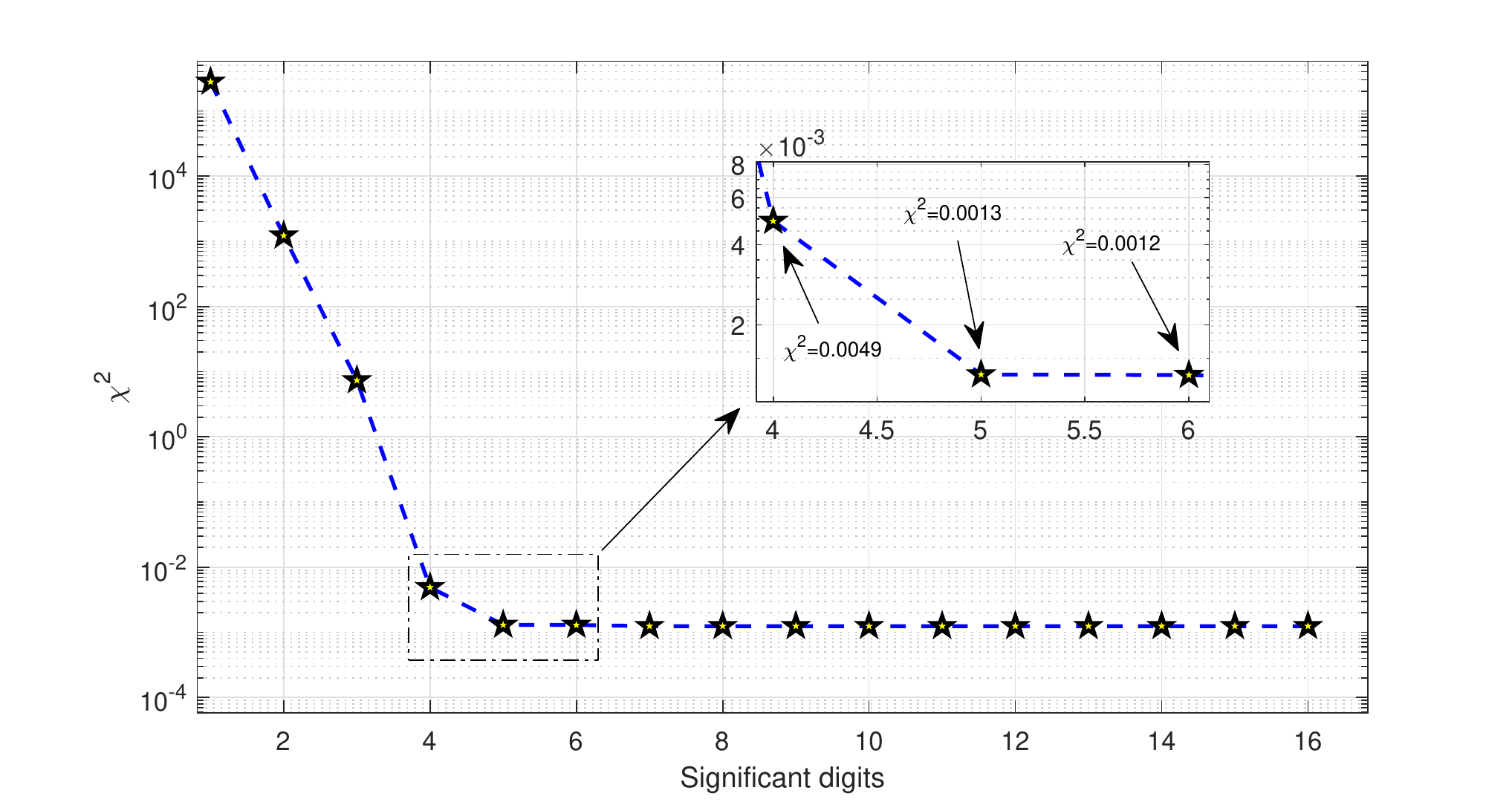}
  \end{center}
  \caption{Plot showing the behavior between the $\chi^2$ value and
    significant digits parameters for scenario $1$.}
  \label{chi2_vs_SD}
\end{figure}

\begin{figure}[htbp]
  \subfloat[\label{subfig:a}]{\includegraphics[scale=0.45]{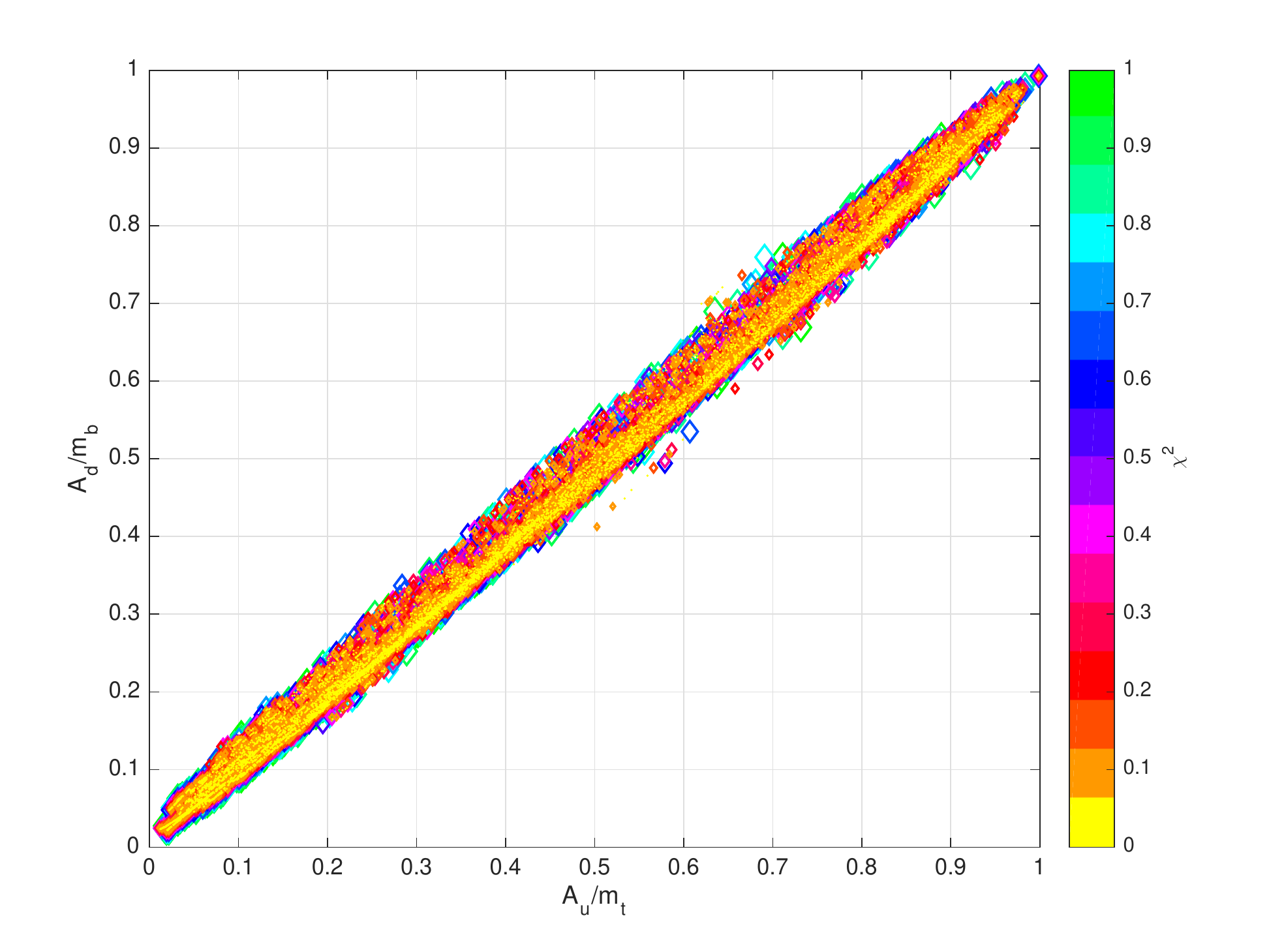}}
  \subfloat[\label{subfig:b}]{ \includegraphics[scale=0.45]{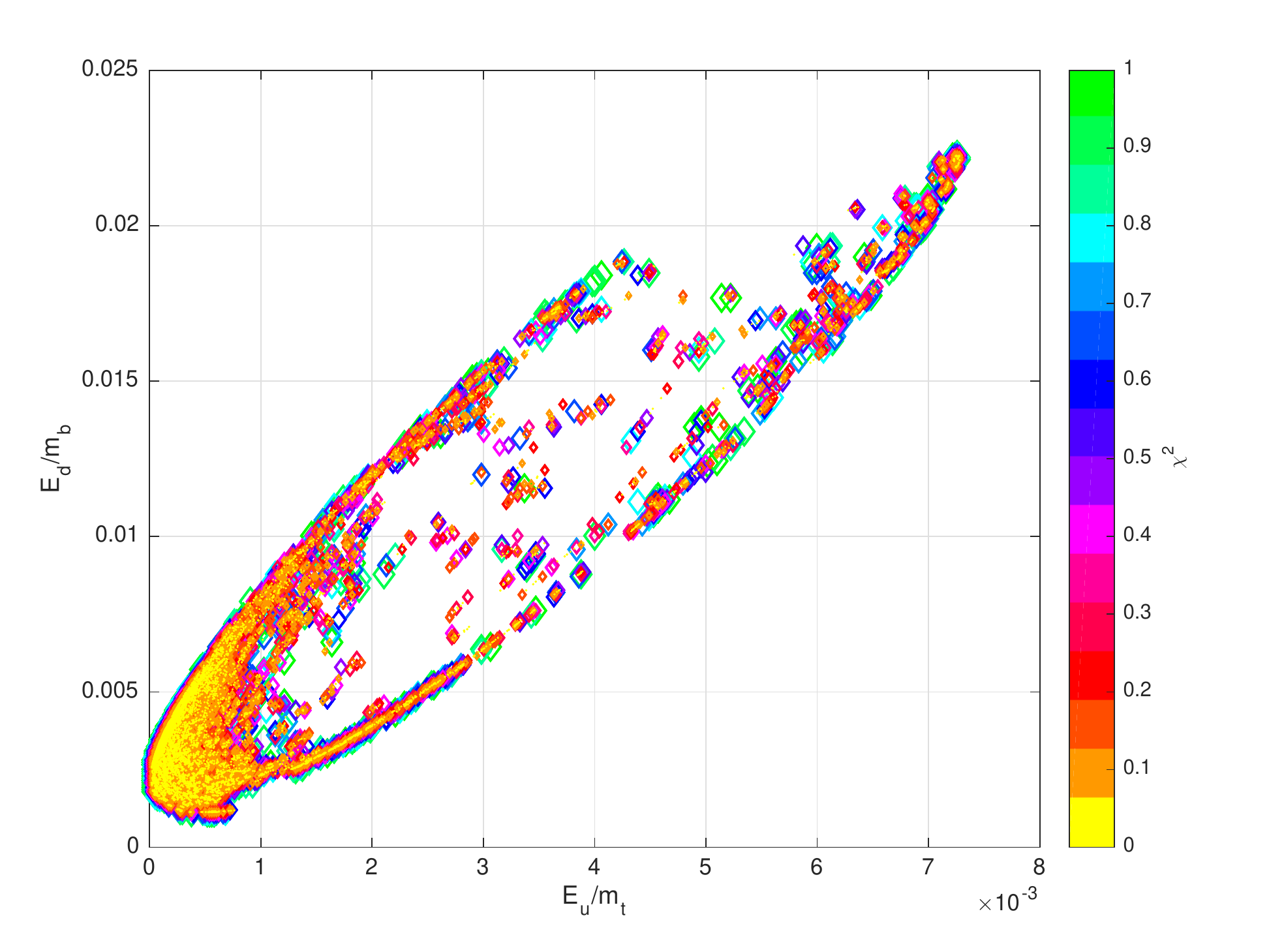}}\hfill\\
  \subfloat[\label{subfig:c}]{\includegraphics[scale=0.45]{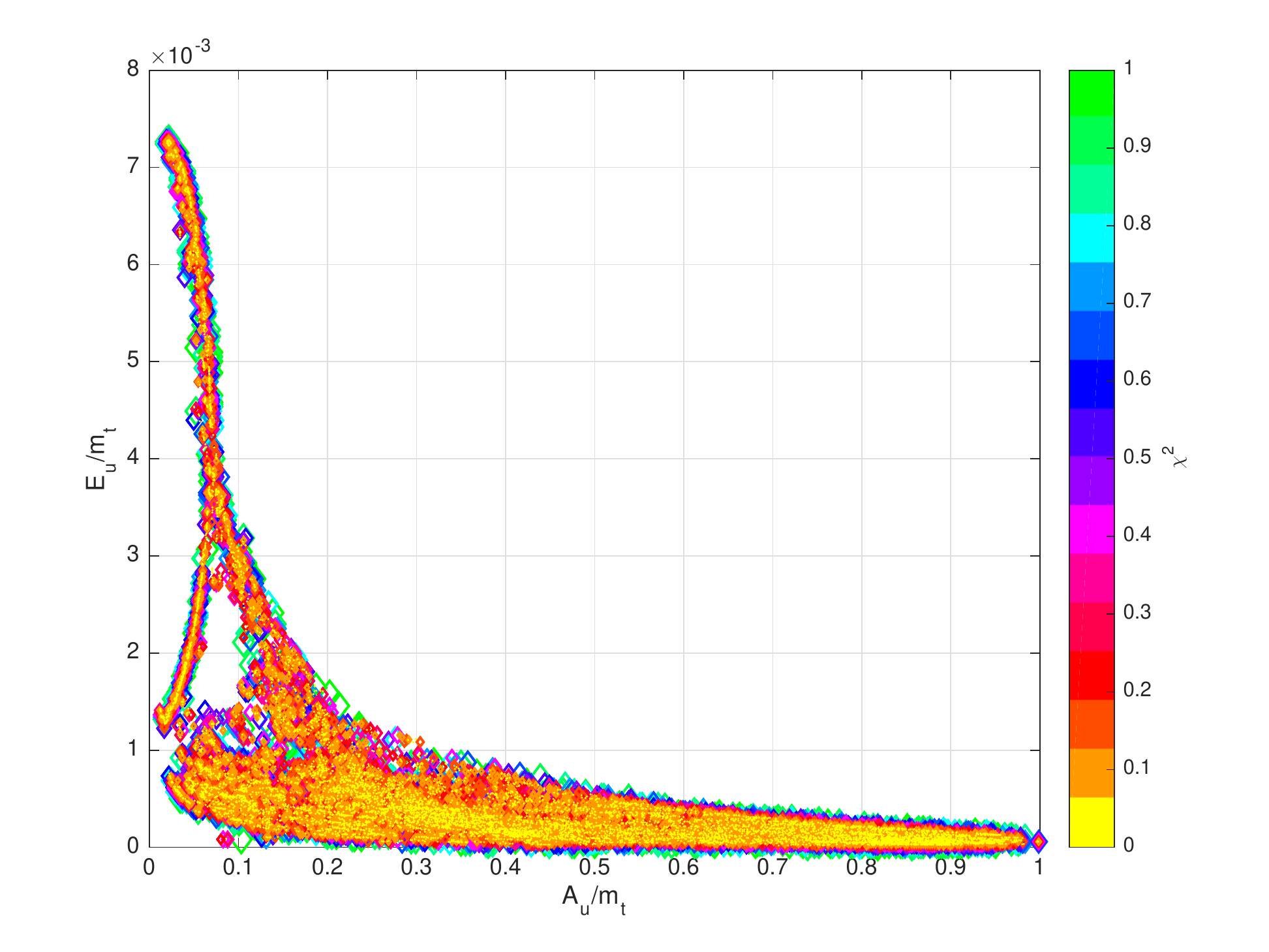}}
  \subfloat[\label{subfig:d}]{\includegraphics[scale=0.45]{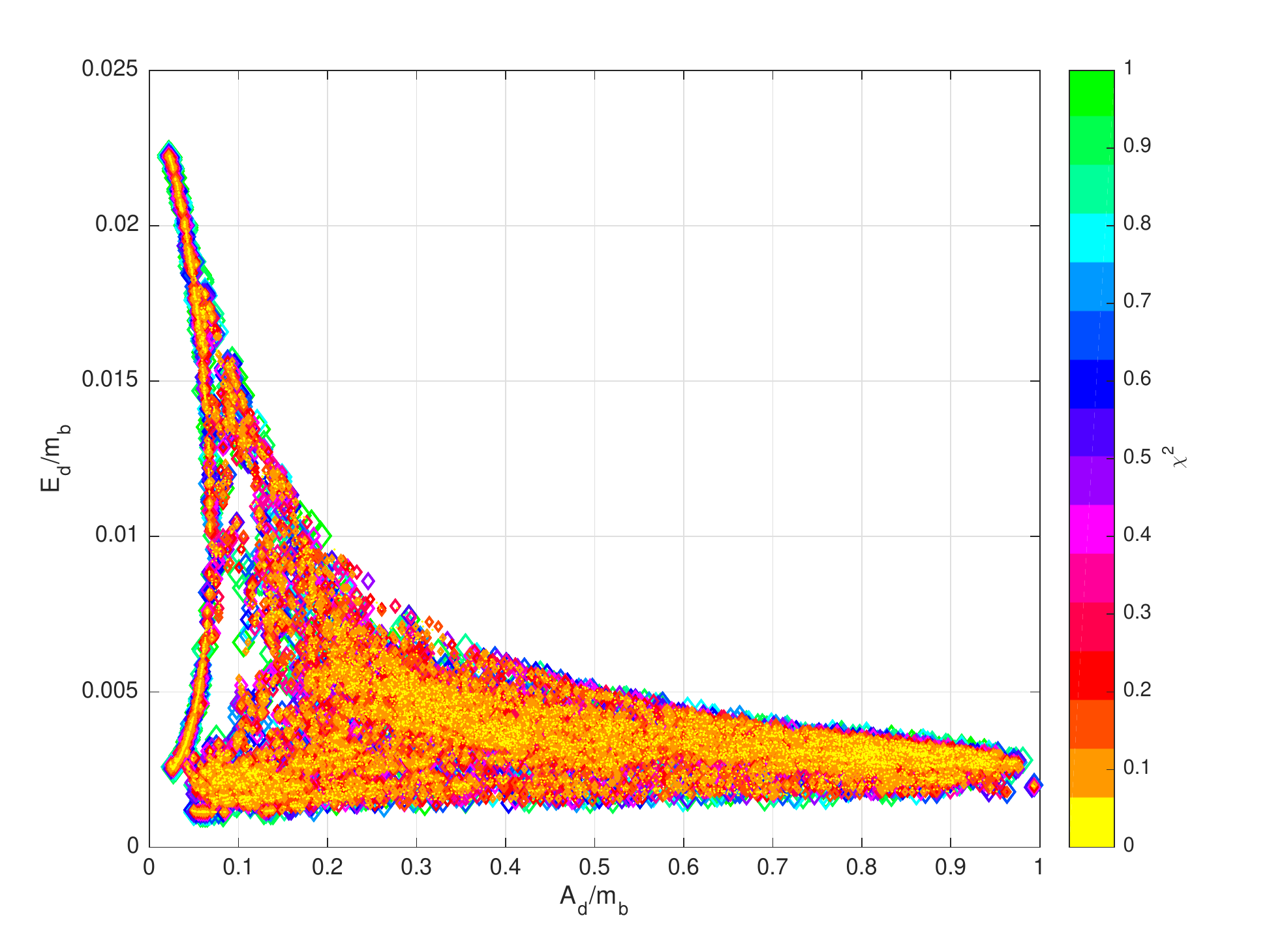}}\hfill
  \caption{In this plot,  the parameter space for the
    scenario 1 of the table \ref{tdatos} is shown, where all parameters are
    positive. Sub figure (a) suggest a linear relation between the
    scaled parameters $A_d/m_{b}$ and $A_u/m_t$ (See further
    discussion in the text). The color bar indicate the values of the
    function in ($\ref{Chi_quarks}$).  }
  \label{fig:parameterspace}
\end{figure}

Among these scenarios (see table \ref{tdatos}) for the quark mixing
matrix in quarks, we have selected the one with real positive free
parameters for in-depth study; this restriction leads to the scenario
1. In any numerical analysis an associated numerical error    
  will appear. We analyze the dependence of $\chi^2$ value with the significant digits in the free parameters. Our result is shown in Figure
  \ref{chi2_vs_SD}, where we can note that, after six significant
  digits the $\chi^2$ is almost constant.  In figure 
\ref{fig:parameterspace}, we show plots of the parameter space
color--coded to show the value of the cost function, and projected on
the $A_u/m_t$ vs. $A_d/m_b$, $E_d/m_b$ vs. $E_u/m_t$, $E_u/m_t$
vs. $A_u/m_t$ and $E_u/m_t$ vs. $A_d/m_b$ planes.
 
A look at subfigure (a), is suggestive of a  linear
  relation between the scaled parameters $A_d/m_b$ and $A_u/m_t$. We compute the Pearson   Coefficient ($\rho$), which has the value $\rho = 0.998$; the linear fit parameters produce a slope $m=0.979$ and a y-intercept $b=0.001$, essentially a straight line at $45^o$.   The correlation between the Yukawa elements $A_u$ and $A_d$ is the form:
$$
\frac{A_d}{m_b} = m\, \frac{A_u}{m_u} + b.
$$
 In subfigure (b) and (d) of Figure \ref{fig:parameterspace}, the
boundaries of the parameter space are clear and the lower bound of
$E_u/m_t$ is compatible with zero. This result implies that the
texture with a zero in the (1,1) place is a possible scenario for the
up-type quark sector. Nevertheless, as it is shown in these plots, the
lower bound of the $E_d/m_b$ parameter never reaches the zero value
for this scenario. This is interesting because it seemingly excludes the 2-zero texture mass matrix for the down type sector.
 
{ A linear correlation between parameters is also present
  in other scenarios. We found a linear behavior between $A_d/m_b$ and
  $A_u/m_t$ for the scenario $3$, and for $E_d/m_b$ and $E_u/m_t$ in
  scenarios $5$ and $7$, with Pearson coefficients $0.996$, $0.986$
  and $0.984$ respectively. }
 
\begin{figure}[htbp]
  \begin{center}
    \includegraphics[scale=0.35]{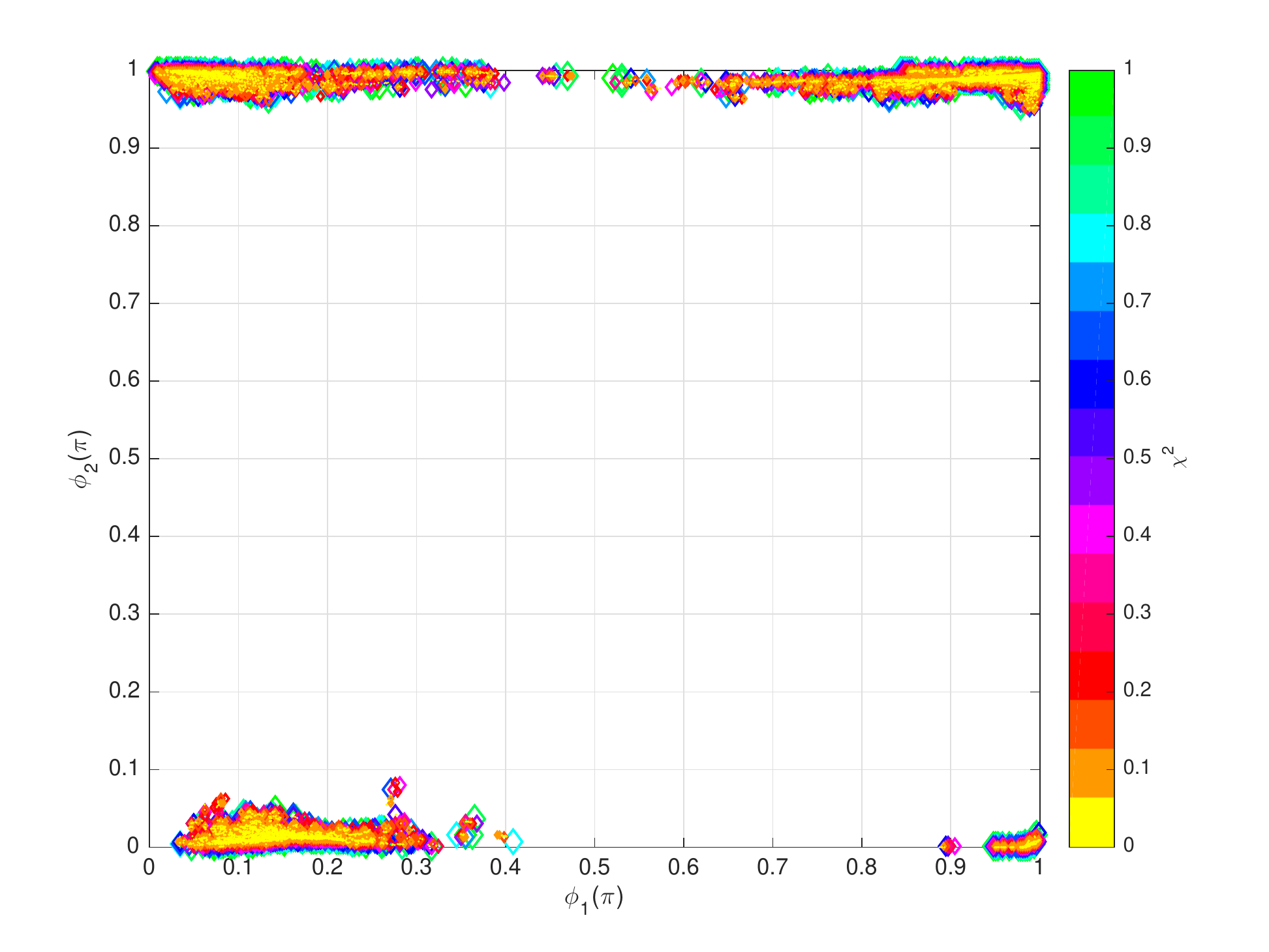}
  \end{center}
  \caption{Plot showing the allowed range of $\phi^q_1$ versus
    $\phi^q_2$, in $\pi$ units. While $\phi_2 \sim 0$ or
    $\phi_2\sim 2 \pi$, the values for $\phi_1$ are not restricted by
    this analysis.}
  \label{phi1_vs_phi2}
\end{figure}

In figure \ref{phi1_vs_phi2}, we have plotted the phases $\phi^q_1$
versus $\phi^q_2$.  The bulk of the points lies on
  the straight lines $\phi^q_2 = 0$, $\phi^q_2 = 2 \pi$ and
  $\phi^q_2 = 4\pi$.  
  $\phi^q_2$. This implies a correlation between the Yukawa
  phases $\phi_{B_u}$ and $\phi_{B_d}$ in the form:
$$
\phi_2 = \phi_{B_u} -\phi_{B_d} = 2n\, \pi,\quad \to \quad \phi_{B_u}
= 2n\, \pi + \phi_{B_d},
$$
where $n$ is an integer number. These two phases are equal in the case
$n=0$.

\section{Numerical analysis of the 2 zero texture in the leptonic
  sector}\label{section5}

\subsection{Comparison between diagonal mass matrix and 1-zero texture
  mass matrix in the charged lepton sector.}\label{section5B}
 
To reduce the number of free parameters, a popular choice of
constraints is a 1-zero neutrino mass matrix in the flavor basis, and
a diagonal mass matrix for the charged leptons
\cite{doi:10.1142/S0217751X13500401}. 

This choice intends to describe all phenomenology of leptonic sector with a reduced number of free parameters. Nevertheless with the advance on the precision of experimental results, such number of free parameters and the restriction imposed by the texture in the neutrino mixing matrix, it is no longer possible to accommodate the updated experimental fittings reported by the collaboration NuFit \cite{Esteban:2018azc}.

\begin{figure}[ht]
  \centering\includegraphics[scale=0.8]{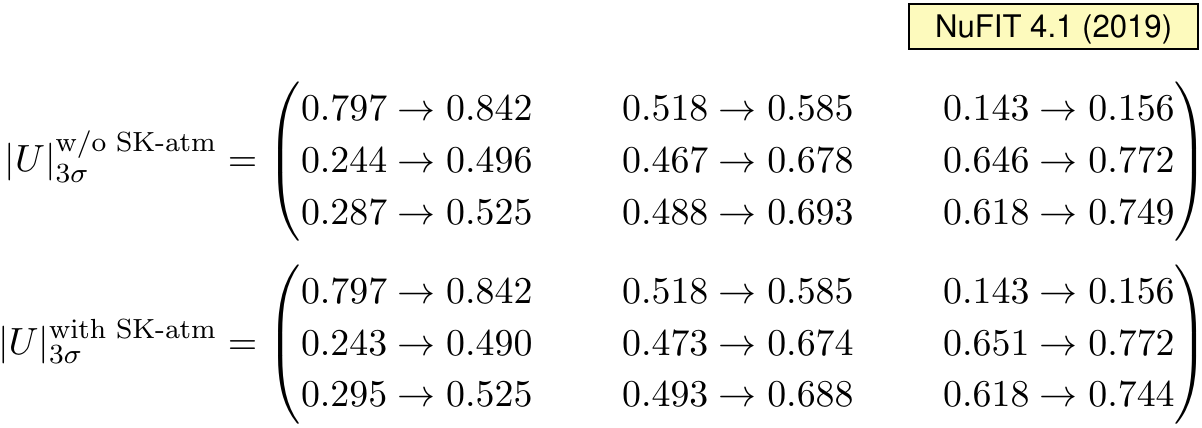}
  \caption{\label{fig:nuFitData} Matrix elements of the PMNS mixing as
    result of the analysis of the group NuFit. Two cases are
    presented, with and without Super Kamiokande experimental
    results. As explained in \cite{Esteban:2018azc}, Super Kamiokande is considered apart from the global fit since there is not enough public information available to reproduce its analysis. Nevertheless, as can be seen in the
      figure, this mainly produces a small difference in the (2,3) entry
      of the mixing matrix. This difference does not drastically change
      the results.}

\end{figure}

 In the parametrization described above, the mass matrix neutrinos can be written as
\begin{equation}\label{eq:massmatrixsimple}
  M^\nu=U_\text{PMNS}\text{Diag}(m_1^\nu,m_2^\nu,m_3^\nu)U_\text{PMNS}^\dagger,
\end{equation}
where the matrix $M^\nu$ is a texture matrix as in
(\ref{1zerotexture}). In order to check the viability of these
results, the fitting in \cite{Esteban:2018azc} was taken for the
elements of the PMNS matrix. These values are shown in figure
\ref{fig:nuFitData}.

In order to obtain the experimentally allowed parameter regions, the
following function is defined as
\begin{equation}\label{eq:chisqsimple}
  \chi^2_{\eta_1\eta_2}=\sum_{i=1}^3 \frac{[U^\text{th}_{ei}-U^\text{exp}_{ei}]^2}{(\Delta U^\text{exp}_{ei})^2}+ \frac{[U^\text{th}_{\mu 3}-U^\text{exp}_{\mu 3}]^2}{(\Delta U^\text{exp}_{\mu 3})^2},
\end{equation}
where the theoretical mixing matrix is
$U_{\alpha i}^\text{th}=U_{\alpha
  i}^\text{th}(A_{\nu},E_{\nu},m_{\nu_\tau},\phi^l_1,\phi^l_2,\eta_1,\eta_2)$
(for $\alpha= e,\mu$).  In the parametrization in use here, the theoretical
  PMNS matrix elements in (\ref{eq:chisqsimple}) are functions of the
  mixing angles and the CP phase; this expressions therefore omits redundant experimental information. 

The $U_{\alpha i}^\text{exp}$ and $\Delta U_{\alpha i}^\text{exp}$ are
the fit from experimental measurement of $U_\text{PMNS}$ matrix
element and the error associated respectively as reported in
\cite{Esteban:2020cvm}. The dependence on the chiral parameters
$\eta_1$ and $\eta_2$ has been taken into account through the
eigenvalues
\begin{align}
  \lambda_1&=\eta_1\sqrt{m_{\nu_3}^2-\Delta m^2_{23}-\Delta m^2_{12}}\label{eq:lambda1}\\
  \lambda_2&=\eta_2\sqrt{m_{\nu_3}^2-\Delta m^2_{23}}\label{eq:lambda2}
\end{align}
with $\Delta m^2_{23}=2.525\times 10^{-3}\text{eV}^2$ and
$\Delta m^2_{12}=7.39\times 10^{-3}\text{eV}^2$. There is a lower
bound from Hermiticity:
$m_{\nu_3}\geq \sqrt{\Delta m^2_{23}+\Delta
  m^2_{12}}$. {As can be seen in the expression
  (\ref{eq:chisqsimple}), the free parameters are restricted 
  using 4 observables, namely $U_{e1}$, $U_{e2}$, $U_{e3}$ and
  $U_{\mu 3}$.  The experimentally allowed region must fulfill $\chi^2_{\eta_1\eta_2}/ N_\text{obs}\leq 1$,
  where $N_\text{obs}=4$. }

The upper bound for $m_{\nu_3}$ was taken from the model-independent
analysis in \cite{PhysRevLett.123.081301} where it is shown that for
Normal Ordering, the sum of neutrino masses fulfills the bound
\begin{equation}
  \sum_i m_{\nu_i}\lesssim 0.26\text{ eV}.
\end{equation}
The definitions of $\Delta m^2_{23}$ and $\Delta m^2_{21}$ allow to
determined the approximated upper bound $m_{\nu_3}\lesssim 0.0959$
eV. In the following, our analysis will be made using this range of
values for the mass of the heaviest neutrino.

It was proceed to compare the behavior of (\ref{eq:chisqsimple}) using two cases; the \emph{diagonal} case represented by the expression (\ref{eq:massmatrixsimple}) and the parallel case where the mixing matrix is written with one zero texture for both  the charged lepton sector and neutrino sector.
 
 The \emph{diagonal} case only consider the free parameters coming from the texture of the neutrino sector and the parallel case increased the number of free parameters. Although this is no the minimal choice, it is consistent with the search of similar flavor structures between the studied sectors.
 
 The analysis was made as follows: first, a random set of parameters was generated, fixing the value for $m_{\nu_3}$ to calculate the function (\ref{eq:chisqsimple}) between the limits imposed by the properties of the mixing matrix and the eigenvalues (\ref{eq:lambda1}-\ref{eq:lambda2}). This process was repeated until obtained a minimal value for the $\chi^2$ . This method was performed for several values of the heaviest neutrino mass.
 
\begin{figure}[ht]
  \begin{center}
    \includegraphics[scale=0.8]{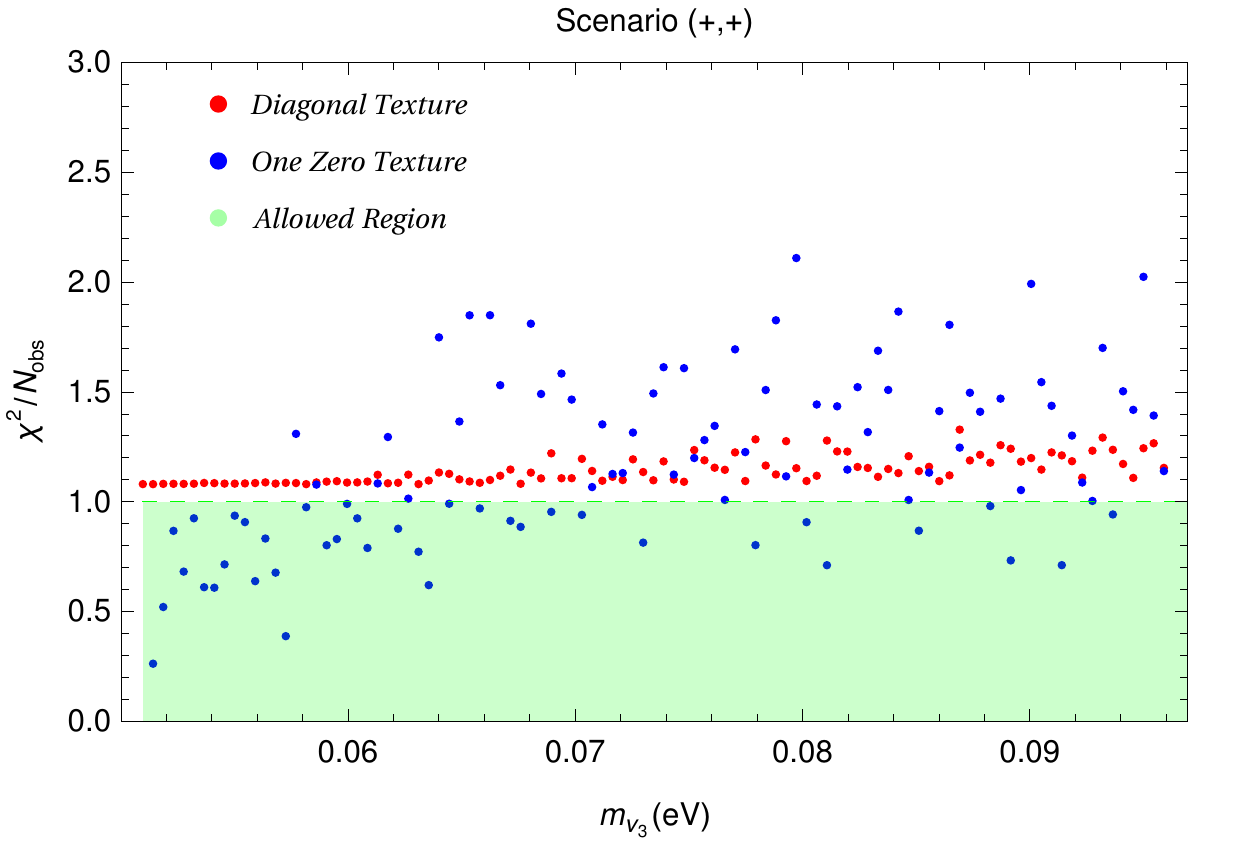}
  \end{center}
  \caption{Behavior of $\chi^2_\text{++}/N_\text{obs}$ with the mass
    of the heaviest neutrino $m_{\nu_3}$ for the case $\eta_1=1$ and
    $\eta_2=1$ using a 1-zero parallel textures and the \emph{diagonal}. The shaded zone shows the allowed experimental region for the $\tilde\chi^2$ function. Every point is the smallest value of the $\tilde\chi^2$ function obtained from a random sample of free parameters. As can be seen, as a consequence of having bigger number of free parameters, the one-zero texture parallel case, have some scenarios allowed compatible with the fitting in \ref{fig:nuFitData}. A diagonal mass matrix in the charged leptonic sector cannot reproduce the experimental fitting of NuFIT 2019.}
  \label{fig:chisqsimple}
\end{figure}

 In figure \ref{fig:chisqsimple}, the plot shows the minimal values for $\chi^2$ with different $m_{\nu_3}$ values for the case with $\eta_1=1$ and $\eta_2=1$. In both cases a 1-zero texture mass matrix was used for the neutrino sector.

With a parallel 1-zero texture, the scaled
$\tilde\chi^2_{\eta_1\eta_2} = \chi^2_{\eta_1\eta_2}/4$ function
depends on 7 continuous parameters. The mixing matrix that comes from
a diagonal mass matrix in the charged lepton sector depends only on
the parameters of the neutrino sector. In order to find a scenario
allowed by the experimental bounds, we searched for a combinations of
free parameters where $\tilde\chi^2_{\eta_1\eta_2}<1$. Evaluating both
cases, it is observed that in the case of a diagonal mass matrix in
the charged lepton sector is not possible to obtain a low enough $\tilde\chi^2$ function. In contrast, with the introduction of the new parameters from the 1-zero mass matrix in the same sector some combination of parameters lying appears in the experimental allowed region.  The cases $(\eta_1,\eta_2)=(-,+),(+,-),(-,-)$ were analyzed, obtaining an allowed region only for the case $(-,+)$ with a similar result as in figure \ref{fig:chisqsimple}.

Once we have numerically demonstrated the viability of the 1-zero
texture mass matrix in the charged lepton and neutrino sectors, in
contrast with the diagonal mass matrix, we test how robust is this
parallel texture independently of the chosen combination of
parameters. 

It was selected the combination of parameters that generates the lowest value for the scaled $\chi^2$ function in both parametrizations, \emph{diagonal} and parallel textures. In order to test the how sensitive is the dependence of the $\chi^2/N_\text{obs}$ on the heaviest neutrinos mass $m_{\nu_3}$ it was varied respect this parameter for both cases.

 The scenarios are characterized by the six parameters
$S_i = (\frac{A_\ell}{m_\tau}, \frac{E_\ell}{m_\tau},
\frac{A_\nu}{m_{\nu_\tau}}, \frac{E_\nu}{m_{\nu_\tau}}, \phi^l_1,
\phi^l_2)$.

To illustrate the behavior, we have chosen the allowed
scenarios in Table \ref{tab:scenarios} representing local minima of
the cost function (\ref{eq:chisqsimple}) calculated numerically.

\begin{table}[ht]
  \begin{tabular}{|c|c|c|}
    \hline
    Scenario & $(\frac{A_\ell}{m_\tau},\frac{A_\ell}{m_\tau},\frac{A_\nu}{m_{\nu_\tau}},\frac{E_\nu}{m_{\nu_\tau}},\phi^l_1,\phi^l_2)$&$\chi^2/N_\text{obs}$\\
    \hline\hline
    Diagonal & $(0.683451,0.0330763,0.25355,0.176368,0.289141,2.19076)$ & $1.07992$\\
    One Zero Parallel & $ (0.439985,0.0530654,0.249244,0.17852,0.564787,0.732588)$ &$0.262339$\\
    \hline
  \end{tabular}
  \caption{Scenarios taken to illustrate the behavior of $\frac{\chi^2_{\eta_1\eta_2}}{N_\text{obs}}$ in terms of
    $m_{\nu_3}$. These are plotted in Figure     \ref{fig:plotchi2}. The prediction for the heaviest neutrino mass depends strongly of the combination of free parameters. In both scenarios are taken to reach its minimum at $m_{\nu_3}= 0.0514286$ eV.}
  \label{tab:scenarios}
\end{table}

\begin{figure}[ht]
  \begin{center}
    \includegraphics[scale=0.8]{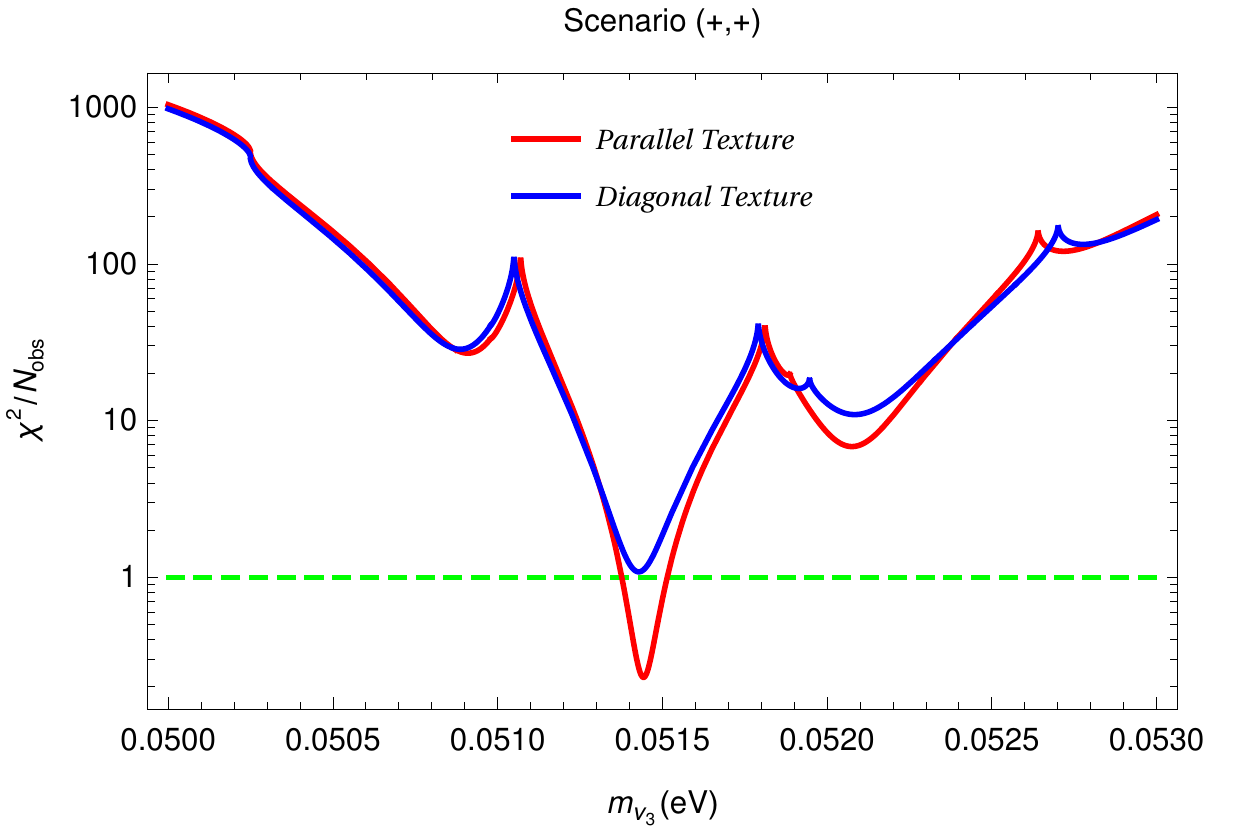}
  \end{center}
  \caption{Scenarios in table \ref{tab:scenarios} that represent the
    local minimum of the $\chi^2$ function in
    (\ref{eq:chi_squared2}). The mass of the heaviest neutrino was
    varied for every scenario. As can be seen in the plot, the minimum
    of the cost function determines different allowed region for the
    $m_{\nu_3}$ values.}
  \label{fig:plotchi2}
\end{figure}

In figure \ref{fig:plotchi2} we plot the
 $\chi^2/N_\textrm{obs}$ function varying $m_{\nu_3}$ around the best fit values for the scenarios described above. As can be seen, the 1-zero texture mass matrix for both sectors, charged leptons and neutrinos, allows a suitable
parameterization for the scale of the heaviest neutrino mass in the
Normal Order case. In the next section, we analyse all possible
parametrizations coming from the parallel texture.

Parallel textures are up to the task of describing the experimental
data; this analysis shows also that present data disfavors a diagonal
shape for the charged lepton mass matrix.  Although this result is
expected due to the introduction of additional parameters, this scenario maintains the idea of the common origin of the mass matrices of the charged lepton and
neutrino sector. Even with the introduction of more parameters in the
parallel case, not all possible theoretical parametrization can
satisfy the updated experimental constraints.

\subsection{Numerical analysis of parallel texture between neutrino
  and charged lepton sector.}\label{sec:B}

In this section, we analyze the case where the mass matrix for the
neutrino sector and charged lepton sector have a parallel 1-zero
texture form. As was mentioned above, the combination of this textures
leads to a mixing matrix with the 2-zero texture. This is the same as
the quark sector case analysed before. 

A complete analysis consist of finding the $8^2$ possibilities to
parameterize the mixing matrix corresponding to the 8 possible
solutions of the system of equations for the invariants. Nevertheless,
in the parallel case where the same mass matrix texture is used in
both sector, charged leptons and neutrinos, only 8 choices are
possible. This is the case that we will study.

A general treatment of the problem starts by finding the PMNS mixing
matrix $U^\text{th}_{\alpha i}$, diagonalizing the matrices
$M_\ell M_\ell^\dagger$ and $M_\nu M_\nu^\dagger$ where
$M_{(\ell,\nu)}$ have a 1-zero texture shape. In total, the mixing
matrix elements are functions on 7 parameters, i.e. $A_{\ell,\nu}$,
$E_{\ell,\nu}$, $m_3^\nu$ and the phases $\phi^l_1$, $\phi^l_2$. In
order to describe the allowed regions in the parameter space, a
$\chi^2$ function was defined as
\begin{equation}\label{eq:chi_squared2}
  \chi^2(\eta_1,\eta_2,\eta_3)=\sum_{i=1}^3 \frac{[U^\text{th}_{ei}(\eta_1,\eta_2,\eta_3)-U^\text{exp}_{ei}]^2}{(\Delta U^\text{exp}_{ei})^2}+ \frac{[U^\text{th}_{\mu 2}(\eta_1,\eta_2,\eta_3)-U^\text{exp}_{\mu 2}]^2}{(\Delta U^\text{exp}_{\mu 2})^2}.
\end{equation}

Because the main goal of this work is to explore the common origin of
the mass matrices of all sectors of the SM, we have restricted the
analysis to the case of Normal Ordering  of the fermion masses
where $m_1^\nu<m_2^\nu<m_3^\nu$. At present, experimental results in
the neutrino sector also allows us to consider Inverse Ordering 
of neutrino masses, with $m_3^\nu<m_1^\nu<m_2^\nu$. The IO of masses
not only changes the sign of the $\Delta m_{32}^2$ observable but also
the theoretical prediction for the $V_\text{PMNS}$ matrix
elements. Although we aim to address this question in a later work, it
can be mentioned that there is a relation between the NO and IO cases
in the neutrino sector through $S_3$ transformations on the neutrino
mass matrix.


The $2^3$ scenarios are characterized by the allowed range of values
for the free parameters and in this section we have established the
notation $E_{\nu,\ell}=e_{\nu,\ell}$ and $A_{\nu,\ell}=a_{\nu,\ell}$
to distinguish these parameters from those analysed in the quark
case. In table \ref{tab:scenarios2}, we show those intervals for the
scenarios defined in the leptonic sector. Some intervals are contained
into others, nevertheless it should be pointed out that in every
scenario, the expressions for the theoretical mixing matrix elements
differs by the combination of the $\eta_i$ parameters defined through
equations (\ref{BF}-\ref{CF}). The masses of the neutrinos are defined
by $m_1^\nu=|\lambda_1|$ and $m_2^\nu=|\lambda_2|$ using the equations
(\ref{eq:lambda1}) and (\ref{eq:lambda2}), leaving the heaviest
neutrino mass $m_{\nu_3}$ as a free parameter. This characteristic
makes the parameter space of the leptonic sector quite different to
the quark sector, where the range of values for the parameters are
fixed by the known quarks masses.

\begin{table}[h]
  \centering
  \begin{tabular}{|c | c|} \hline
    Scenario & Intervals \\
    \hline
    1: & $-m_3^F<e_F<-m_2^F \wedge -m_2^F<a_F<-m_1^F$ \\
    2: & $-m_3^F<e_F<-m_2^F \wedge -m_2^F<a_F<m_1^F$ \\
    3: & $-m_3^F<e_F<-m_1^F \wedge -m_1^F<a_F<m_2^F$ \\
    4: & $-m_3^F<e_F<m_1^F \wedge m_1^F<a_F<m_2^F$ \\
    5: & $-m_2^F<e_F<-m_1^F \wedge -m_1^F<a_F<m_3^F$ \\
    6: & $-m_2^F<e_F<m_1^F \wedge m_1^F<a_F<m_3^F$ \\
    7: & $-m_1^F<e_F<m_2^F \wedge m_2^F<a_F<m_3^F$ \\
    8: & $m_1^F<e_F<m_2^F \wedge m_2^F<a_F<m_3^F$ \\
    \hline
  \end{tabular}
  \caption{Interval of values for $a_\nu$, $a_\ell$, $e_\nu$ and
    $e_\ell$ in different scenarios for the charged lepton and
    neutrino sectors. Has been taken Normal Order for the neutrino
    masses as the charged lepton masses, this is $m_1^F<m_2^F<m_3^F$
    with $F=\ell,\nu$.}
  \label{tab:scenarios2}
\end{table}

To select a region of allowed parameters, the
$\chi^2$ function was compared to the number of non-correlated observables. In this
analysis those were 4, i. e. $U_{e1}$, $U_{e2}$,$U_{e3}$ and
$U_{\mu 1}$. The interval of possible values for $a_\nu$ and $e_\nu$
depends on the observables $\Delta m^2_{32}$, $\Delta m^2_{21}$ and
$m^\nu_3$.

\begin{figure}[htbp]
  \begin{center}
    \includegraphics[scale=0.7]{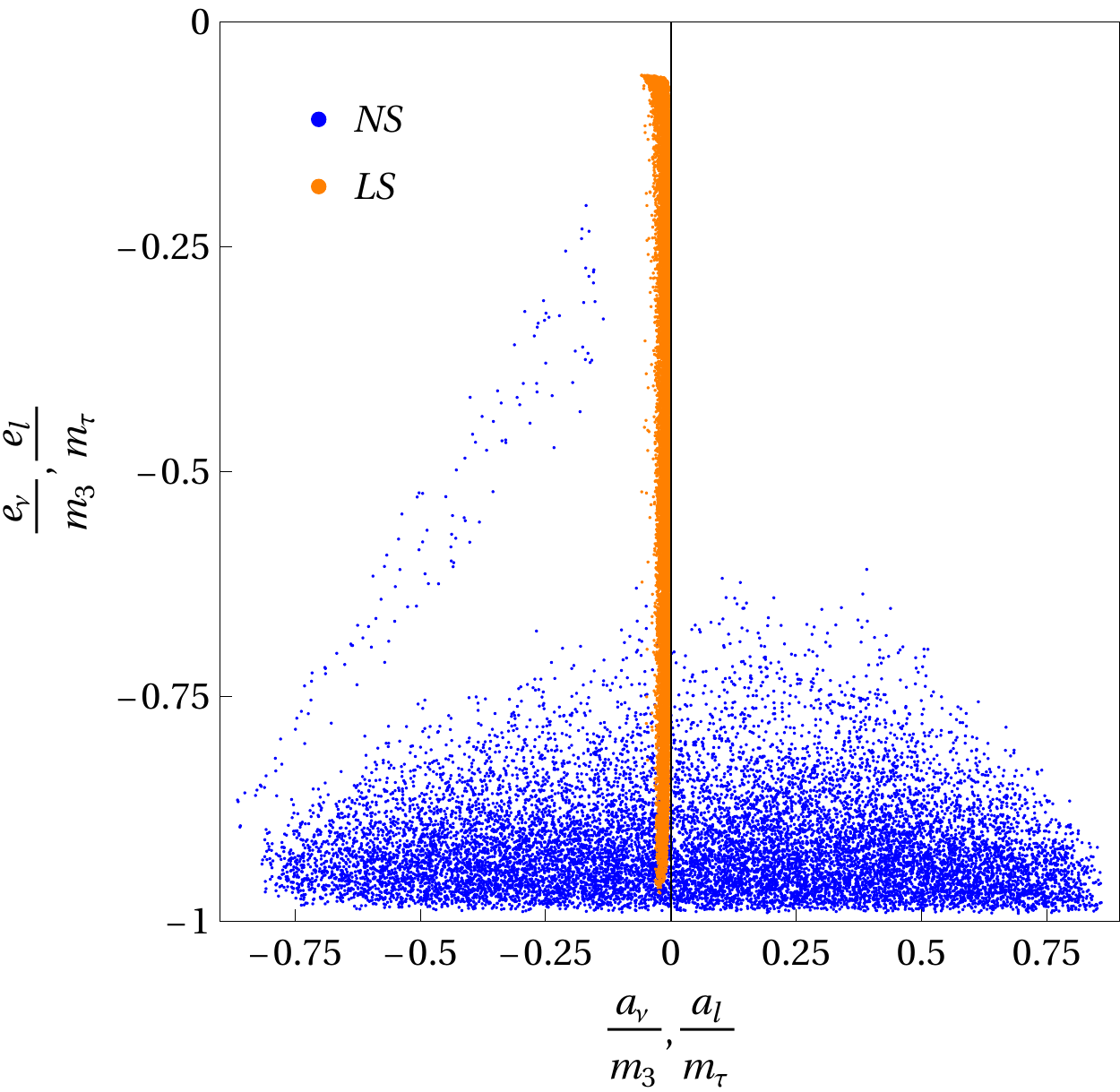}
  \end{center}
  \caption{\label{SCPlotsSc2} In this plot it is shown the scaled
    scatter plot for $e_\nu/m_{\nu_3} \text{ vs } a_\nu/m_{\nu_3}$
    Neutrino Sector (NS) in the scenario 2, with random values for the other
    parameters. The limits of the plot are determined by the values of
    the table \ref{tab:scenarios2} calculated for a random value of
    the heaviest neutrino mass fulfilling the experimental
    bounds. Also it is shown the projection on the space
    $e_{\ell}/m_\tau$ vs $a_\ell/m_\tau$ Leptonic Sector (LS). The interval of the
    parameters in the charged lepton sector are fixed and scaled with
    $m_\tau$. As can be seen in the plot, there is a narrow common
    region where the values for the LS  are similar
    to the NS}
\end{figure}

\begin{figure}[htbp]
  \begin{center}
    \includegraphics[scale=0.7]{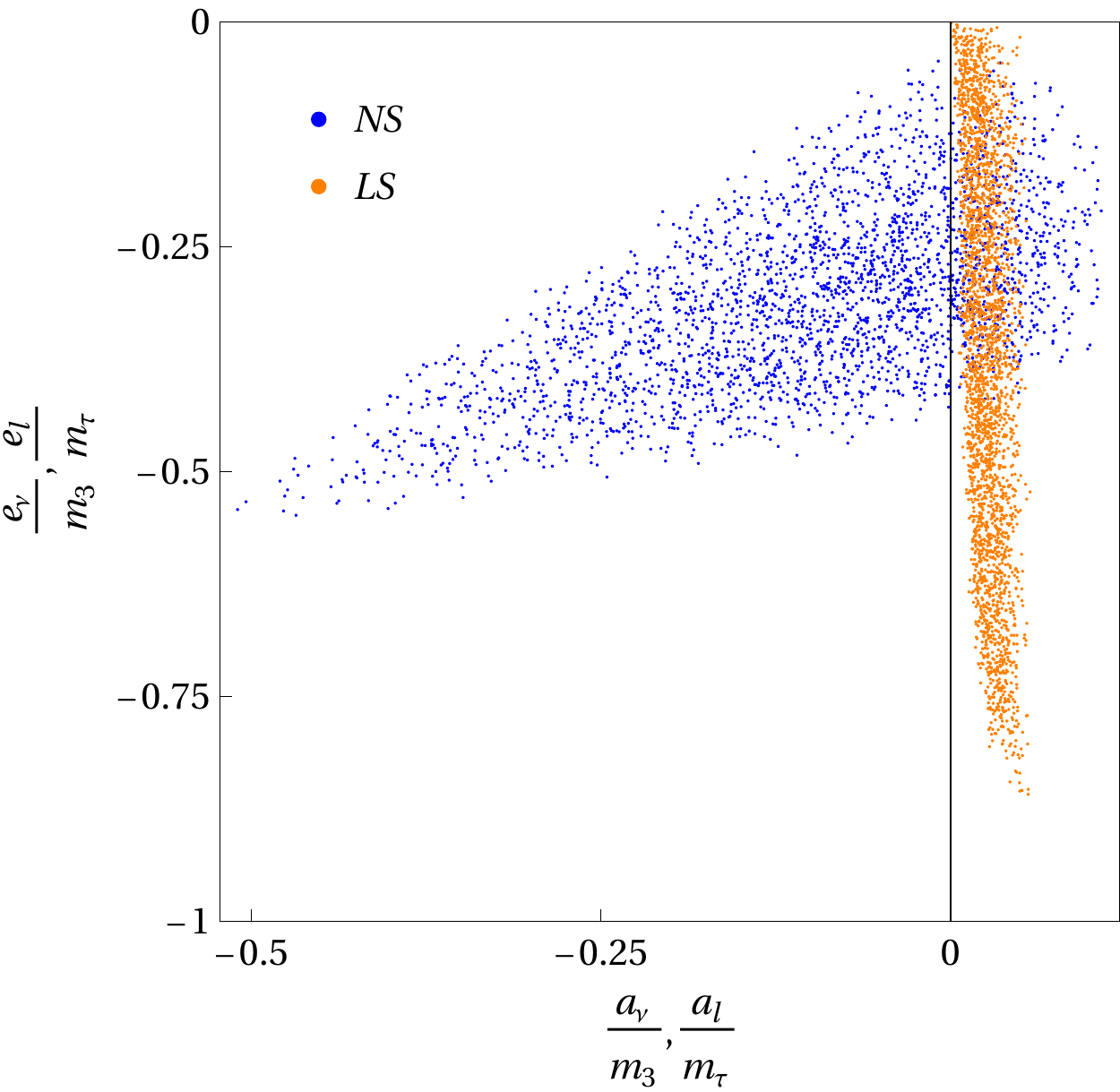}
  \end{center}
  \caption{\label{SCPlotsSc3} In this plot we show the behavior of the
    experimentally allowed values in the scenario 3 for the
    projections $e_\nu/m_{\nu_3} \text{ vs } a_\nu/m_{\nu_3}$ (NS) and
    $e_\ell/m_{\tau} \text{ vs } a_\ell/m_{\tau}$ (LS), using random
    values for the other parameters. The region corresponding to the
    parameters for the LS have a common region with
    the obtained for the NS. This region in on the
    positive values for the $a$'s parameters in both sector and the
    $-0.5\lesssim \frac{e_{\nu,(\ell)}}{m_3(m_\tau)}\lesssim 0$
    region}
\end{figure}

\begin{figure}[htbp]
  \begin{center}
    \includegraphics[scale=0.65]{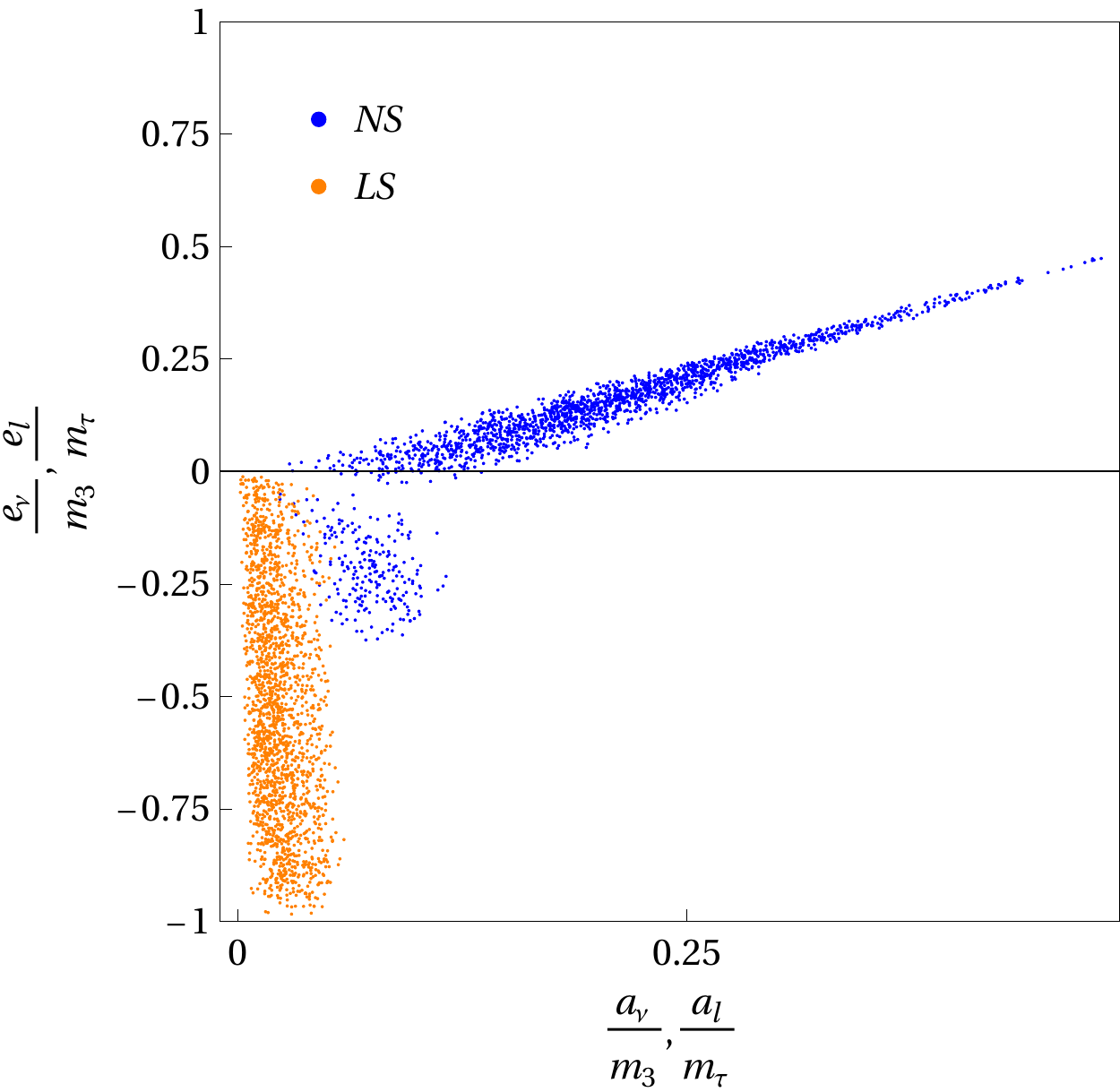}
  \end{center}
  \caption{\label{SCPlotsSc4} Allowed points corresponding to the
    scenario 4 for $e_\nu \text{ vs } a_n$ using random values for the
    other parameters. The matching region between scaled parameters in
    the charged lepton and neutrino sector is below the horizontal
    line $e_\nu/m_{\nu_3}=e_\ell/m_\tau=0$.}
\end{figure}

\begin{figure}[htbp]
  \begin{center}
    \includegraphics[scale=0.65]{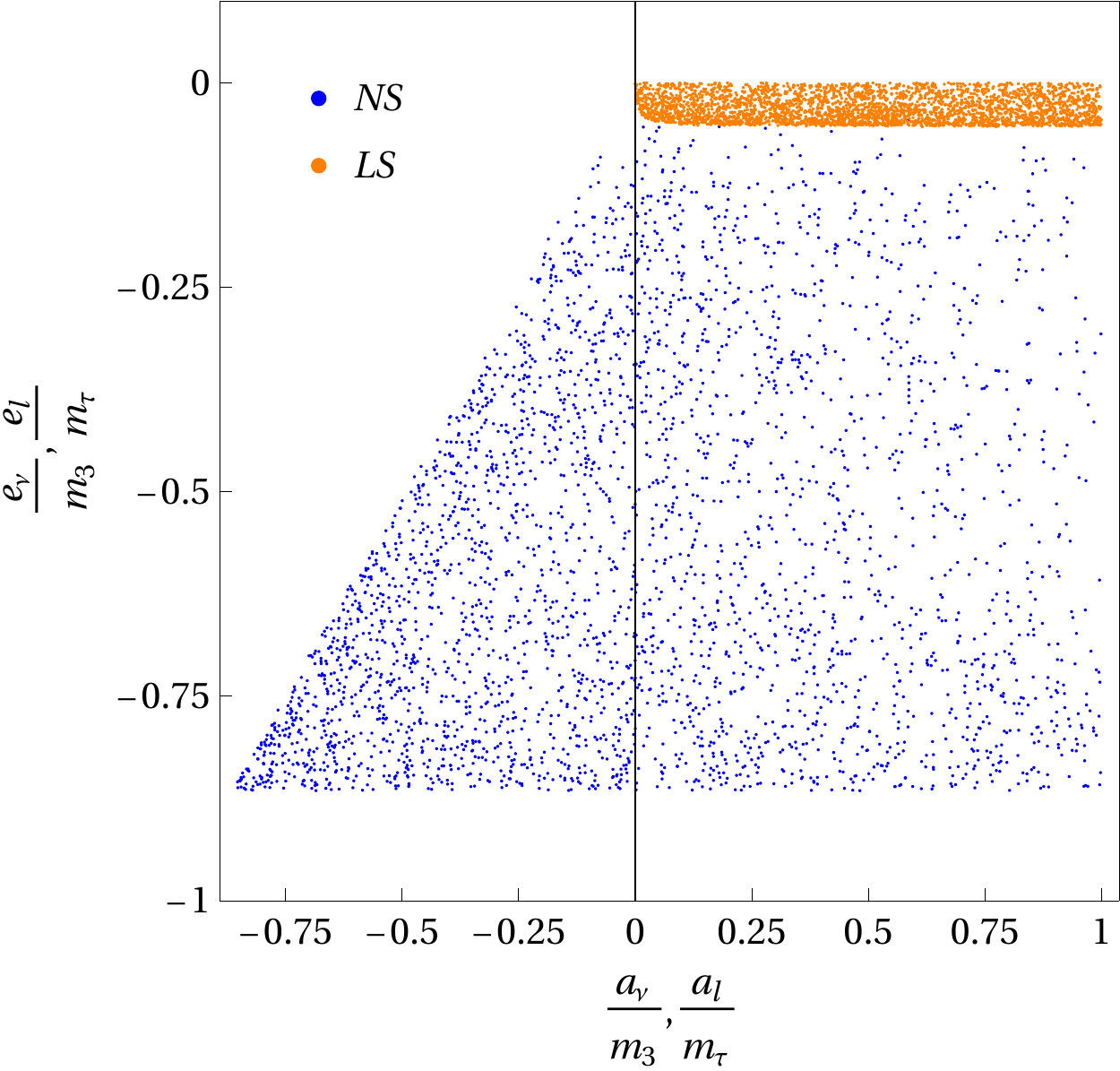}
  \end{center}
  \caption{\label{SCPlotsSc5} Allowed points corresponding to the
    scenario 5 for $e_\nu \text{ vs } a_n$ using random values for the
    other parameters. The matching region between scaled parameters in
    the charged lepton and neutrino sector is in the right-hand side
    of the vertical line $a_\nu/m_{\nu_3}=0$ in the figure of the
    left.}
\end{figure}

\begin{figure}[htbp]
  \begin{center}
    \includegraphics[scale=0.6]{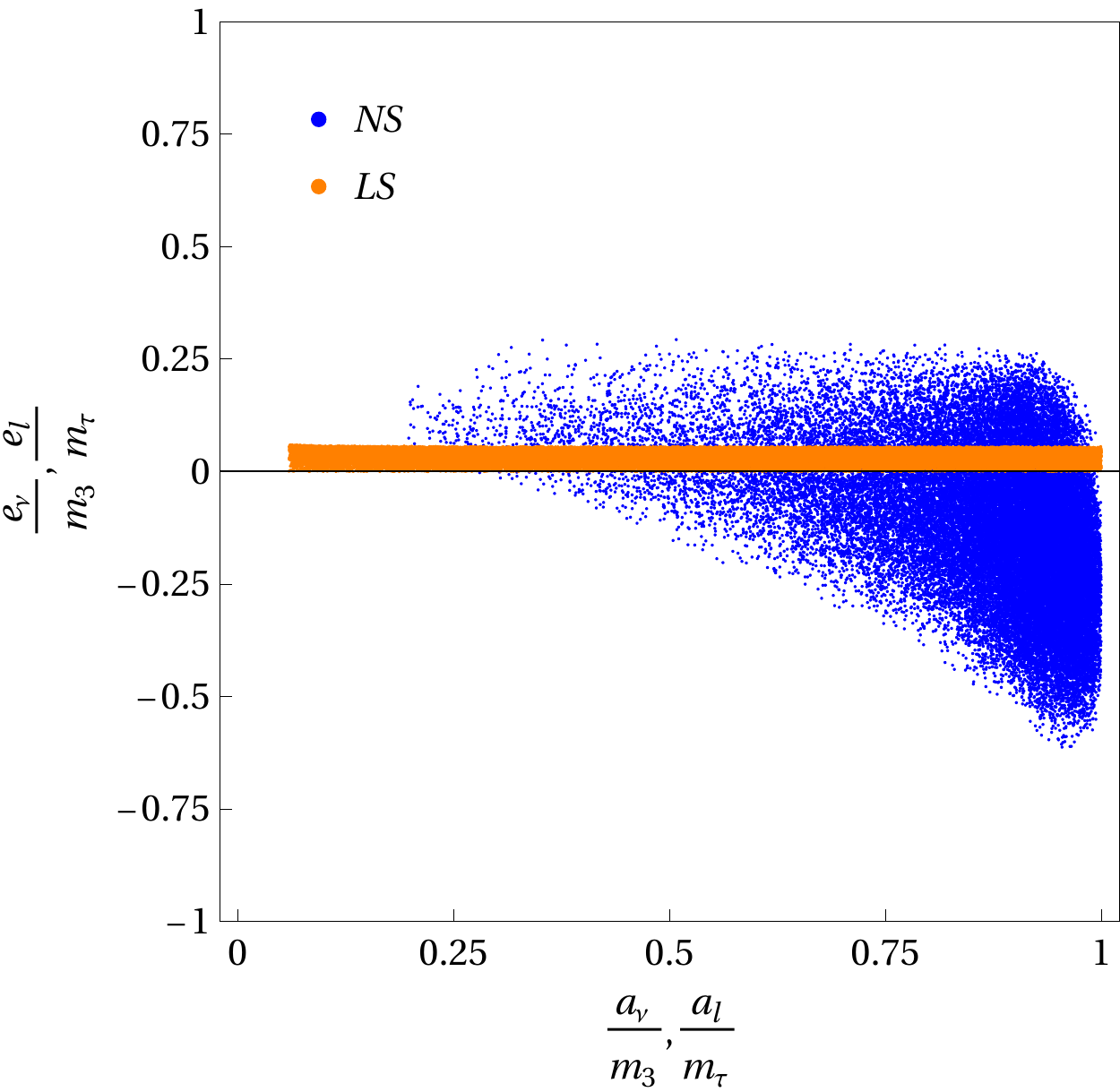}
  \end{center}
  \caption{\label{SCPlotsSc7} Allowed points corresponding to the
    scenario 7 for $e_\nu \text{ vs } a_n$ space using random values
    for the other parameters. The matching region between scaled
    parameters in the charged lepton and neutrino sector is in the
    upper side of the horizontal line $e_\nu/m_{\nu_3}=0$ in the
    figure of the left.}
\end{figure}

\begin{figure}[htbp]
  \begin{center}
    \includegraphics[scale=0.6]{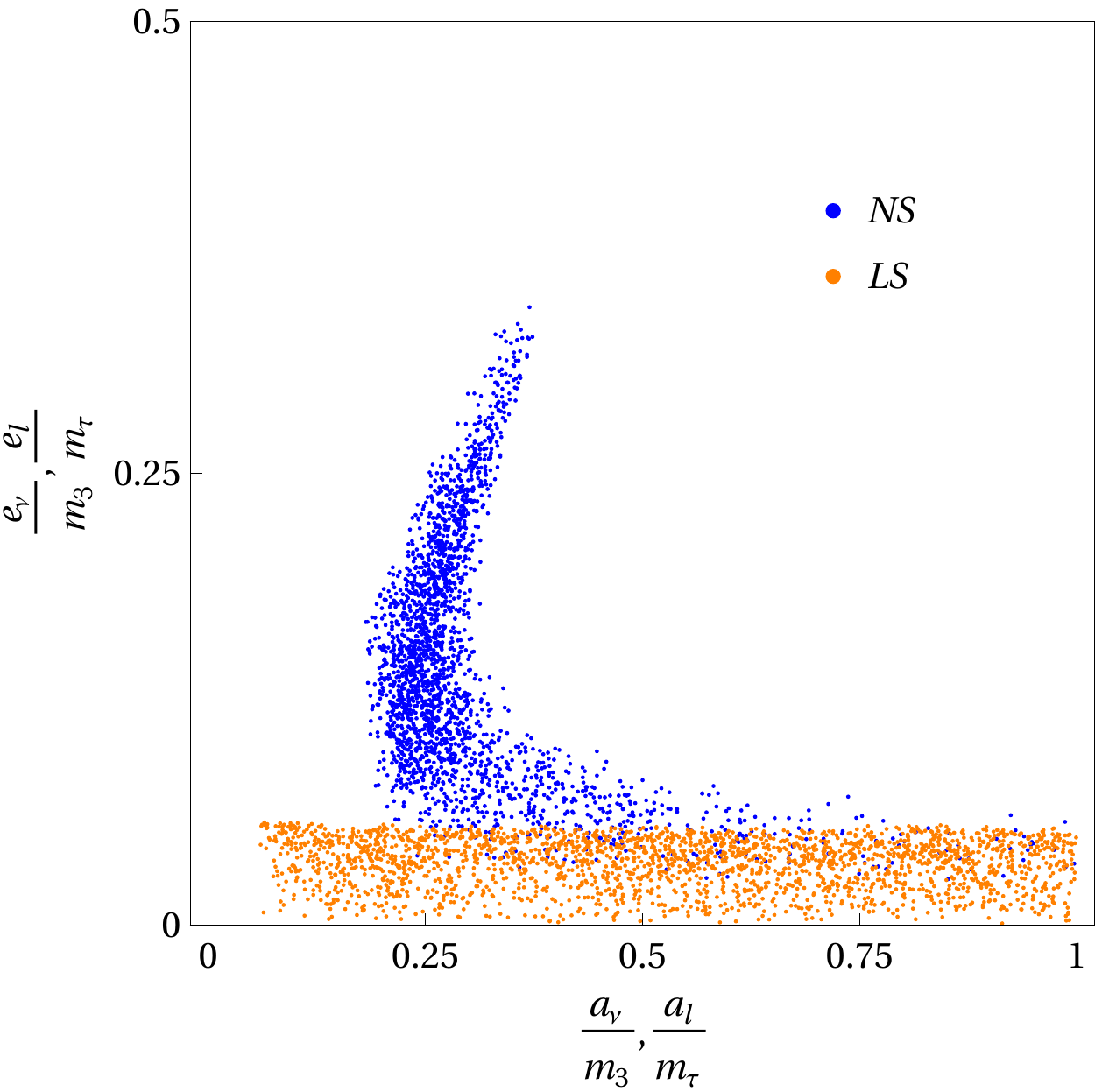}
  \end{center}
  \caption{\label{SCPlotsSc8} Allowed points corresponding to the
    scenario 8 with all parameters real positive, for
    $e_\nu \text{ vs } a_n$ and using random values for the other
    parameters.}
\end{figure}

As can be seen in the figures (\ref{SCPlotsSc2}- 
\ref{SCPlotsSc8}), some region have clear boundaries. The shape
and extension of these regions are determined by the specific form
of the function (\ref{eq:chi_squared2}) and the interval for the 
parameters in the table \ref{tab:scenarios2}. Samples to 
generate points in all scenarios were calculated with the same 
statistical conditions, i. e., the number of points in the plot 
are proportional to the probability of having a combination of 
values compatible with the chosen observables. Thus the 
boundaries of the region represent the
combination of restrictions for the intervals in table
\ref{tab:scenarios2} with the experimental observables.

The main result of this analysis is that scenarios 1 and 6 are discarded because 
the restriction on free parameters can not generate $\chi^2$ 
values smaller than one. In figures (\ref{SCPlotsSc2}) to 
(\ref{SCPlotsSc8}), we present scatter plots of the parameter 
space for the experimentally allowed scenarios. The fitting used 
to find the data were taken from figure \ref{fig:nuFitData}.

Figure \ref{SCPlotsSc2} shows points that represent experimentally
allowed parameter values for the scenario 2. It shows scaled
parameters with respect to the heaviest fermion mass for every sector,
and leads to similar values in a narrow region determined mainly by
the restriction on $\frac{a_\ell}{m_\tau}$. As a consequence we can
explore the possibility of a Universal Texture Constraint (UTC) as
described in \cite{Carrillo-Monteverde:2020fie} for the Yukawa sector
in extended scalar models, but in this case for the SM. The UTC was
introduced in \cite{Carrillo-Monteverde:2018eqi} to reduce drastically
the number of free parameters in models with parallel textures.

As in scenario 2, it is also possible to find a matching region in the
scenario 3 where the values of the scaled parameters can be equal. For
this scenario 3, the matching region lies in the right hand side of
the vertical line at $a_\nu/m_{\nu_3}=0$ (see the plot in Figure
\ref{SCPlotsSc3}). The matching region is thus
$-0.5\lesssim \frac{e_{\nu,(\ell)}}{m_3(m_\tau)}\lesssim 0$ and
positive values for $a_\ell/m_{\tau}$.

Similar observations can be made for the scenarios 4, 5, 7 and 8 shown
in figures \ref{SCPlotsSc4}, \ref{SCPlotsSc5}, \ref{SCPlotsSc7} and
\ref{SCPlotsSc8}, respectively. Interestingly, in the scenario 8 (as
seen in Figure \ref{SCPlotsSc8}), the allowed region for the neutrino
sector is completely contained in the allowed region of the charged
lepton sector. In the following section, we analyze this matching to
support the possible presence of an UTC.

A possible common origin of the mass matrices in leptonic and quark
sectors is not evident in the mixing matrix because the mixing angles
are very different, and the masses span a wide range of order of
magnitude.

If there is a relation between these sectors it could be better
manifested in scaled parameters over the heaviest fermion masses,
which we endeavor to study next.

\section{The scenario with positive parameters in the leptonic
  sector}\label{section6}
In this section, we study the scenario where $a_\nu$, $a_\ell$,
$e_\nu$ and $e_\ell$ are all positive to find the region of free
parameters that can reproduce both the neutrino and the charged lepton
texture compatible with the mixing angles and the CP phase. To use the
same scheme as in the quark sector, we have only considered NO of the neutrino masses.
 
  As can be seen in the Figure \ref{subfig:a},the results suggest a
  linear relationship between the scaled parameters $A_d/m_b$ and
  $A_u/m_t$. This is in the scenario when all parameters are positive,
  and for this reason we have chosen the same case but in the leptonic
  sector in order to look for similar behaviors in the parameters. 
  
 \begin{figure}[htbp]
   \centering \includegraphics[scale=0.65]{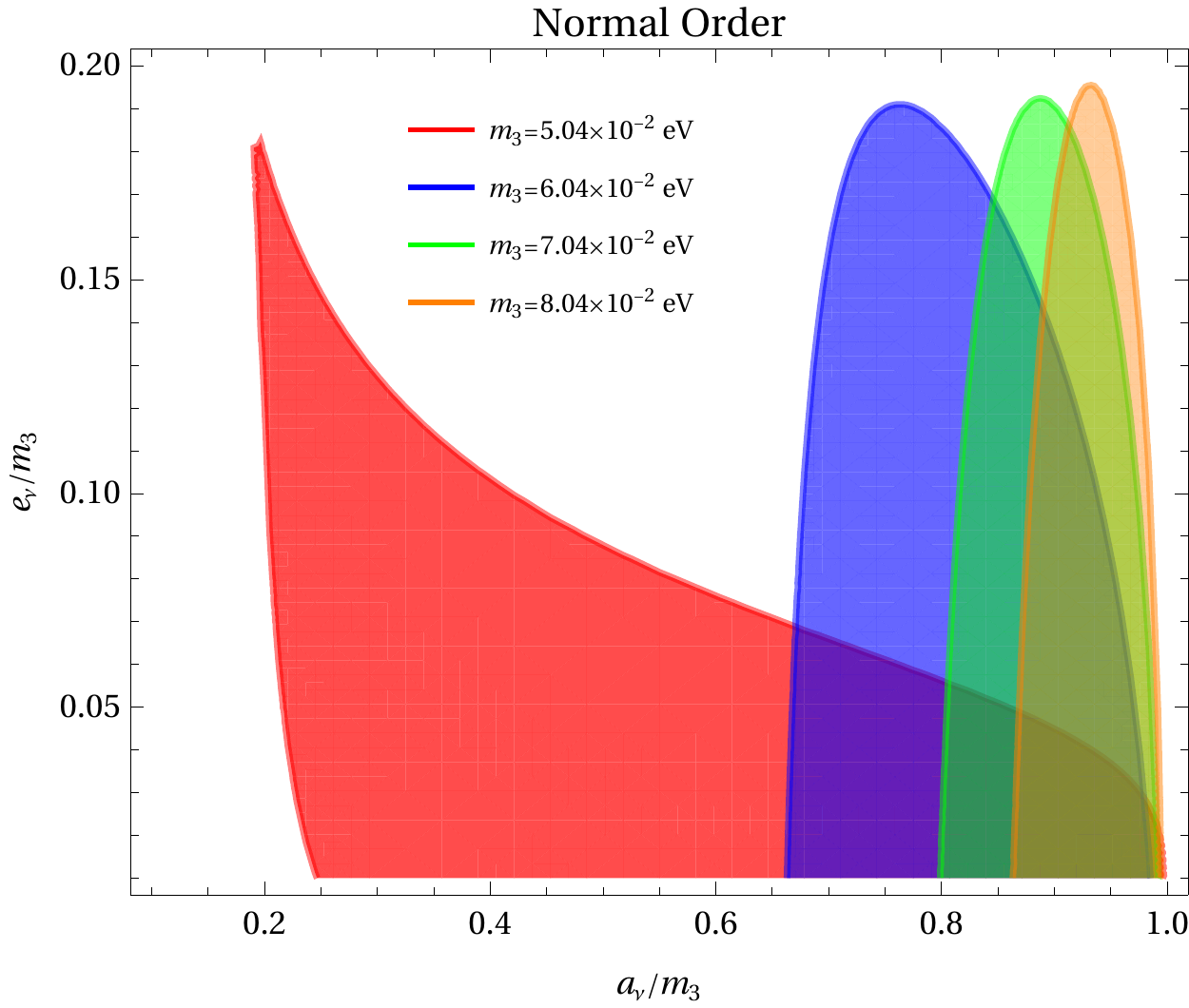}
   \caption{Dependence on the heaviest mass neutrino $m_{\nu_3}$ of
     the allowed regions in the $a_\nu/m_{\nu_3}$ vs
     $e_{\nu}/m_{\nu_3}$ parameter space. This plot was generated
     using the values $a_\ell/m_\tau=0.597$, $e_\ell/m_\tau=0.05094$,
     $\phi_1\simeq 1.86\pi$ and $\phi_2\simeq1.65\pi$}
   \label{fig:SCPlotsAnEn}
 \end{figure}

 In order to determine the behavior of the regions of the
 $a_\nu/m_{\nu_3}$ vs $e_{\nu}/m_{\nu_3}$ parameter space in the
 leptonic sector with respect to the heaviest neutrino mass, we
 selected the minimum point for the $\tilde\chi^2$ function. Then,
 analytical regions in the $a_\ell/m_{\tau}$ vs $e_{\ell}/m_{\tau}$
 for several values of $m_{\nu_3}$ were calculated while using the
 values previously determined for other parameters. These results are
 summarized in figure \ref{fig:SCPlotsAnEn} where it is noticed how
 with the increase of $m_{\nu_3}$, the interval of possible values for
 $a_\nu/m_{\nu_3}$ narrows down. On the other hand, the range for
 $e_\nu/m_{\nu_3}$ remains approximately constant. Increasing the
 values for $m_{\nu_3}$ out of the range shown in the plot reduces the
 allowed region until it disappears. An experimental measurement of
 the mass $m_{\nu_3}$ out of the range generating the allowed regions
 would clearly discard this scenario.

\begin{figure}[htbp]
  \centering \includegraphics[scale=0.6]{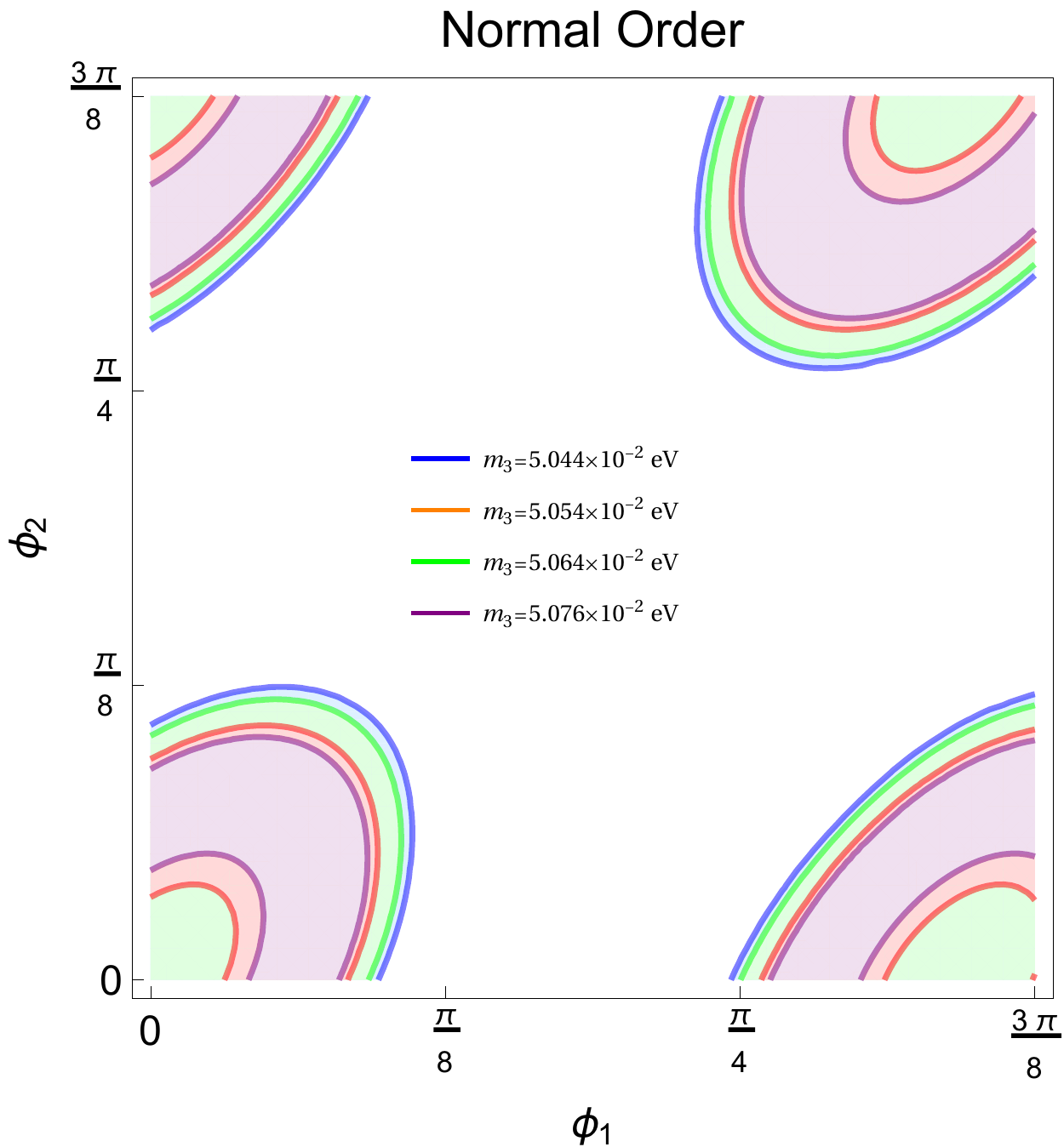}
  \caption{Allowed regions in the phase space obtained varying the
    mass $m_{\nu_3}$. To obtain this regions the parameters
    $a_\ell/m_\tau$, $e_\ell/m_\tau$, $a_\ell/m_\tau$ and
    $e_\ell/m_\tau=$ were fixed by the $\chi^2$ minimum.}
  \label{fig:SCPlotph1ph2}
\end{figure}

Using the same methodology, we analysed the behavior of the allowed
regions in the phases $\phi_1$ vs $\phi_2$. The allowed region is very
sensitive to the mass $m_{\nu_3}$ as presented in Figure
\ref{fig:SCPlotph1ph2}.

\begin{figure}[htbp]
  \centering \includegraphics[scale=0.65]{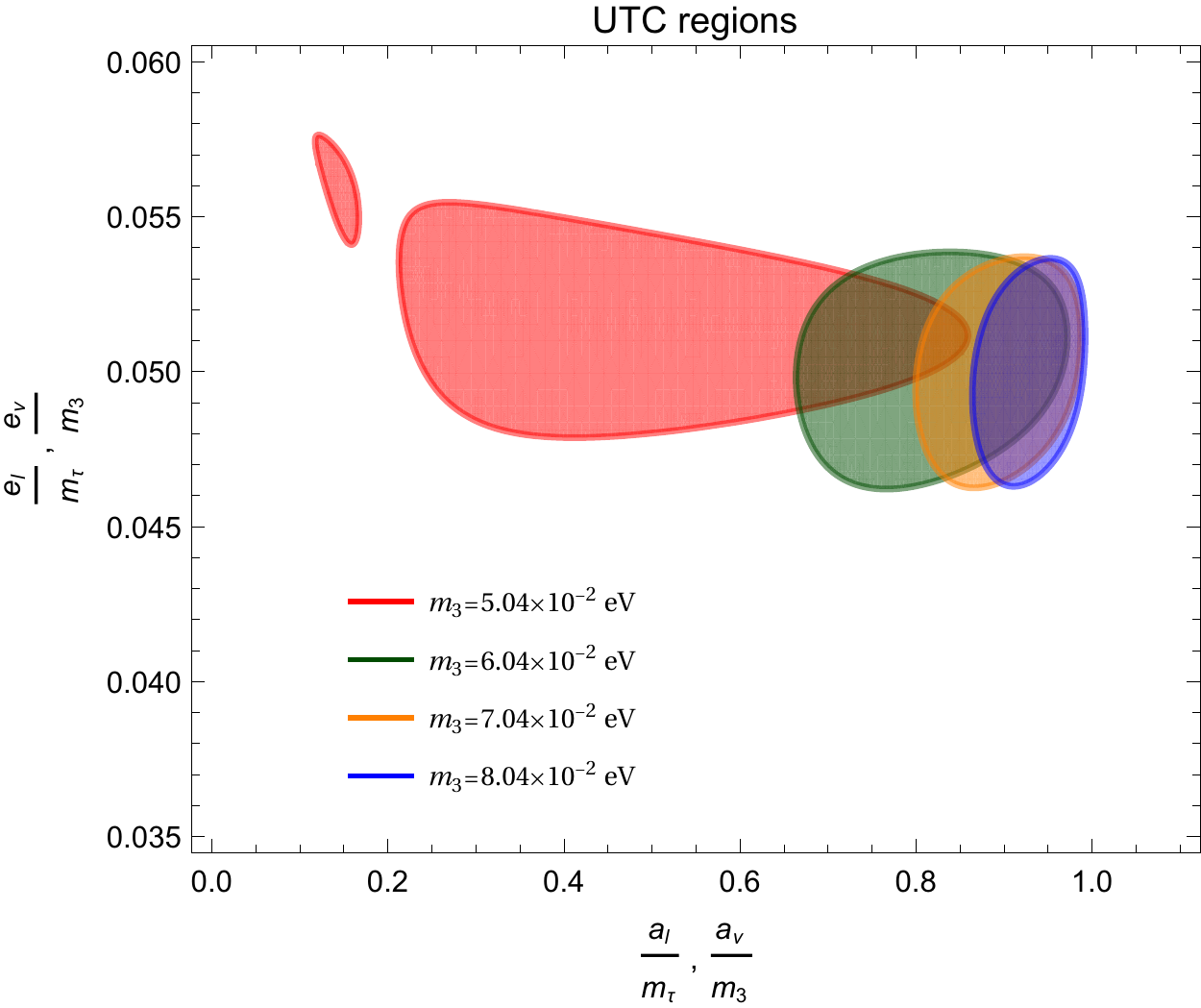}
  \caption{In this plot, the region of parameters allowed
    by the experimental fitting using Universal Texture Constraint  it is shown for $m_{\nu_3}=5.0442\times 10^{-2}$eV,
    $m_{\nu_3}=6.0442\times 10^{-2}$eV,
    $m_{\nu_3}=7.0442\times 10^{-2}$eV and
    $m_{\nu_3}=5.0442\times 10^{-2}$eV.}
  \label{fig:SCPlotCombined2}
\end{figure}

In order to evaluate the validity of the UTC in the parallel texture,
in Figure \ref{fig:SCPlotCombined2} we present together the allowed
regions using the condition $a_\ell/m_\tau=a_\nu/m_{\nu_3}$ and
$e_\ell/m_\tau=e_\nu/m_{\nu_3}$ for different values of the heaviest
neutrino mass.

The allowed region gets narrow and small when the heaviest neutrino
mass approaches the experimental bound mentioned in section
\ref{section5B}. This result suggests that the 1-zero texture with UTC
makes the parallel texture parametrization a more predictive model for
the leptonic sector. It is worth to mention that a radical application
of the UTC can relate the lepton and quark sectors of the SM shedding
light on an underlying flavor physics beyond SM.

\section{Conclusions}\label{section7}

We have obtained exact expressions for the VCKM and PMNS matrix
elements in terms of the free parameters in the scenario of a 1-zero
parallel texture for all leptons and quarks.

Numerical work was done using a hybridized nature-inspired/cellular
automata search algorithm CPSO-DE, supplemented with constraints. We
studied the limit where some parameters go to zero, to compare with
previous texture matrices used in the literature. This analysis
concludes that narrowing experimental constraints are in tension with
the diagonal plus 1-zero texture case studied in \citep{liu2013} and
\citep{Verma:2010jy}. It also shows that the zero-parameter limit for
type-down quarks, when the 1-zero texture reduces to the two-zero
texture, is excluded.

Similar to the UTC discussed in
\citep{Carrillo-Monteverde:2020fie}, we find that leptons and quarks
can be described by the same structure. Eight scenarios, corresponding
to the eight possibilities in (\ref{eq:chi_squared2}), were
considered.  In six of those, as detailed above, parameter values
consistent with all experimental constraints were found, and the
minima for the cost function reported; for two scenarios, the local
minima found are excluded by this criteria.

Even though this model requires an enlarged parameter space, it
provides in exchange a common framework for the description of leptons
and quarks. Furthermore, the explored regions for the parameters
describing the quark mass matrices suggest that, once scaled, some
parameters for up-type and down-type quarks are approximately equal.
A reduction on the number of parameters can be achieved by imposing
some sort of universality constraint. For the positive--parameters
case studied, there is a narrow region where the same values of the
charged leptons and neutrinos parameters (with Dirac masses and Normal
Ordering) provide reasonable agreement with experimental data. This is
a more extreme version of a UTC, with a significant parametric
reduction.

\begin{acknowledgments}
  This work has been supported in part by\textit{ SNI-CONACYT
    (M\'exico)}. The authors thank A. Carrillo-Monteverde for reading
  the manuscript and for her valuable comments.
\end{acknowledgments}

\bibliographystyle{apsrev4-2}

\bibliography{main}
\end{document}